\documentclass{aa}
\usepackage{graphicx}
\usepackage{txfonts}
\usepackage{natbib,twoopt}
\usepackage[breaklinks=true]{hyperref} 
\usepackage{xcolor}

\begin{document}

\title{Astrophysical properties of star clusters in the Magellanic Clouds
homogeneously estimated by \texttt{ASteCA}}
\subtitle{}

   \author{G.I. Perren
          \inst{1,3}\thanks{\email{gabrielperren@gmail.com}}
          \and
          A.E. Piatti
          \inst{2,3}
          \and
          R.A. V\'azquez\inst{1,3}
          }

   \institute{Facultad de Ciencias Astron\'omicas y Geof\'{\i}sicas (UNLP),
              IALP-CONICET, La Plata, Argentina
        \and
        Observatorio Astron\'omico, Universidad Nacional de C\'ordoba,
        C\'ordoba, Argentina
        \and
        Consejo Nacional de Investigaciones Cient\'{\i}ficas y T\'ecnicas
        (CONICET), Buenos Aires, Argentina\\
    }

   \date{Received Aug 8, 2016; accepted Jan 20, 2017}

 
\abstract
{}
{To produce an homogeneous catalog of astrophysical parameters of 239 resolved
star clusters located in the Small and Large Magellanic Clouds, observed in the
Washington photometric system.}
{The cluster sample was processed with the recently introduced Automated
Stellar Cluster Analysis (\texttt{ASteCA}) package, which ensures both an
automatized and a fully reproducible treatment, together with a statistically
based analysis of their fundamental parameters and associated uncertainties.
The fundamental parameters determined with this tool for each cluster, via a
color-magnitude diagram (CMD) analysis, are: metallicity, age, reddening,
distance modulus, and total mass.}
{We generated an homogeneous catalog of structural and fundamental parameters
for the studied cluster sample, and performed a detailed internal error analysis
along with a thorough comparison with values taken from twenty-six published
articles.
We studied the distribution of cluster fundamental parameters in both Clouds,
and obtained their age-metallicity relationships.}
{The \texttt{ASteCA} package can be applied to an unsupervised determination of
fundamental cluster parameters; a task of increasing relevance as more data
becomes available through upcoming surveys.}


\maketitle
%

\section{Introduction}
\label{sec:intro}

The study of a galaxy's structure, dynamics, star formation history, chemical
enrichment history, etc., can be conducted from the analysis of its star
clusters.
Star clusters in the Magellanic Clouds (MCs) are made up of a varying number of
coeval stars sharing a chemical composition, assumed to be located
relatively at the same distance from the Sun, and affected by roughly the same
amount of reddening. These factors facilitate the estimation of
their fundamental parameters, and thus the properties of their host galaxy.
New developments in astrophysical software allow the homogeneous
processing of different types of star clusters' databases. The article series
by Kharchenko et al. 
\citep[see][and references therein]{Kharchenko_2005,Schmeja_2014}
and the integrated photometry based derivation of age and mass for 920 clusters
presented
in~\cite{Popescu_2012}, based on their MASSsive CLuster Evolution and
ANalysis package~\citep[MASSCLEAN,][]{Popescu_2009}\footnote{
\url{http://www.massclean.org/}}, are examples of semi-automated and automated
packages applied on a large number of clusters.

However, there is no guarantee that by employing a homogeneous method, we will
obtain similar parameter values from the same cluster photometric
data set across different studies. This is particularly true when the methods
require user intervention, which makes the results subjective to some degree.
In~\cite{Netopil_2015} the open cluster parameters age, reddening, and distance
are contrasted throughout seven published databases. The authors found that all
articles show non-negligible offsets in their fundamental parameter values.
This result highlights an important issue: most of the color-magnitude diagram 
(CMD) isochrone fits are done by-eye, adjusting correlated parameters
independently, and often omitting a proper error treatment \citep[see]
[for a more detailed description of this problem]{vonHippel_2014}.
When statistical methods are employed, the used code is seldom publicly
shared to allow scrutiny by the community. There is then no objective way to
asses the underlying reliability of each set of results.
Lacking this basic audit, the decision of which database values to use
becomes a matter of preference.

As demonstrated by~\cite{Hills_2015}, assigning precise fundamental
parameters for an observed star cluster is not a straightforward task.
Using combinations of up to eight filters ($UBVRIJHK_s$) and three stellar
evolutionary models, they analyzed the NGC188 open cluster with a
Bayesian isochrone fitting technique implemented in their 
BASE-9 package\footnote{\url{http://webfac.db.erau.edu/~vonhippt/base9/}}, and
arrived at statistically different results depending on the isochrones and the
filters used.
NGC188 is a ${\sim}4$ Gyr cluster with a well defined Main Sequence (MS) mostly
unaffected by field star contamination, and with proper motions and radial
velocities data available.
A typical situation -- where a star cluster is observed through fewer
filters, affected by a non-negligible amount of field star contamination, and
without information about its dynamics -- will be significantly more
complicated to analyze.
A mismatch between theoretical evolutionary models, along with an
inability to reproduce clusters in the unevolved MS domain, had already
been reported in~\cite{Grocholski_2003}.

The aforementioned difficulties in the analysis of star cluster CMDs will only
increase if the study is done by-eye, since: a) the number of possible solutions
manually fitted is several orders of magnitude smaller than that handled by a
code, b) correlations between parameters are almost entirely disregarded, c)
uncertainties can not be assigned through valid mathematical
means -- and are often not assigned at \mbox{all --}, and d) the final values
are necessarily highly subjective.
In the last thirty years, many authors have applied some form of
statistical analysis to derive star clusters' fundamental parameters.
We give in Sect 2.9 of~\citet[][hereafter Paper I]{Perren_2015} a non-exhaustive
list of articles where these methods were employed.
Still, by-eye studies continue to be used. This is because most statistical
methods developed are either closed-source, or restricted to a particular form
of analysis (or both).
The need is clear for an automated general method with a fully open
and extensible code base, that takes as much information into account as
possible, and capable of generating reliable results.\\

In Paper I we presented the Automated STEllar Cluster Analysis
(\texttt{ASteCA}) package, aimed at allowing an accurate and comprehensive study
of star clusters.
The code is released under a GPL v3 general public
license\footnote{\url{https://www.gnu.org/copyleft/gpl.html}}, and
can be downloaded from its official
site\footnote{\url{http://asteca.github.io}}.
Through a mostly unassisted process the code analyses clusters' positional
and photometric data sets to derive their fundamental parameters and
uncertainties.
As shown in Paper I, the code is able to assign precise parameter values for
clusters with low to medium field star contamination, and gives reasonable
estimations for heavily contaminated clusters. Every part of this astrophysical
package is open and publicly available, and its development is ongoing.

In the present work we apply \texttt{ASteCA} on 239 clusters in the Small and
Large Magellanic Clouds (S/LMC), distributed up to ${\sim}5^{\circ}$ and $
{\sim}8^{\circ}$ in angular distance from their centers, respectively.
The MCs are located close enough to us to allow the study of their resolved star
clusters. The large number of cataloged clusters -- $\sim4000$ are listed in
the~\cite{Bica_2008} catalog -- makes them an invaluable resource for
investigating the properties of the two most massive galaxies that orbit the
Milky Way.
The reddening that affects the MCs is relatively small, except for a few regions
like 30 Doradus in the LMC, where $E_{B-V}$ can reach values above $0.4$
mag~\citep{Piatti_2015b}. The overall low levels of reddening simplifies
the research of the clusters in these two galaxies.
We use photometric data sets in the $CT_1$ the Washington
system~\citep{Canterna_1976,Geisler_1996}, known to be highly sensitive
to metal abundance for star clusters older than ${\sim}1$
Gyr~\citep{Geisler_1999}.
The results obtained here regarding the metal content, are thus of relevance for
the analysis of the MCs chemical enrichment history.

This is the first study where such a large sample of resolved star clusters is
homogeneously analyzed in an automatic way, with their fundamental
parameters statistically estimated rather than fitted by eye or fixed a priori.
Having metal content assigned for 100\% of our sample is particularly
important, especially compared to other star cluster catalogs. The latest
version of the well known DAML02 database~\citep[v3.5, 2016 Jan 28;][]
{Dias_2002}\footnote{\url{http://www.wilton.unifei.edu.br/ocdb/}},
for example, reports abundances for only 13\% out of the 2167
clusters cataloged. Estimations for the total cluster mass is given in few
cases, if integrated photometry is provided.

This article is structured as follows.
In Sect.~\ref{sec:clust-sample} we present the star cluster sample used in
this work along with numerous studies used to compare and validate our
results.
Sections~\ref{sec:fund-params} and~\ref{sec:errors-fit} describe the estimation
of the fundamental parameters derived with \texttt{ASteCA}, and analyze their
uncertainties, respectively.
In Sect.~\ref{sec:comp-pub-data} a detailed comparison of our results with
published values from the literature is performed.
Sect.~\ref{sec:param-dist} shows the distribution of cluster fundamental
parameters in our catalog, and the age-metallicity relationships (AMRs) for the
cluster system.
Sect.~\ref{sec:summ-concl} summarizes our results and concluding remarks.


\section{Clusters sample}
\label{sec:clust-sample}

\begin{figure}
\centering
\includegraphics[width=\hsize]{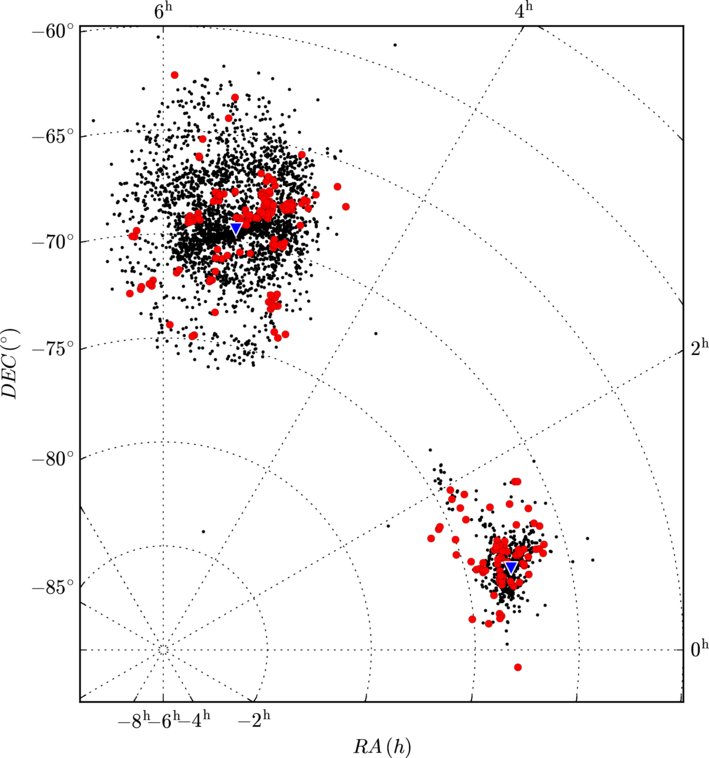}
\caption{Distribution of our set of analyzed clusters (red circles)
superimposed on to the~\cite{Bica_2008} database of 3740 star clusters (black
dots), for both MCs. The assumed centers for the Clouds are marked with blue
triangles.}
\label{fig:ra-dec}
\end{figure}

The data set used in this work consists of $CT_1$ Washington photometry for 239
star clusters; 150 of them located in the LMC and the rest in the SMC.\@
These clusters were selected because they were readily available, are
already analyzed in the literature -- meaning we can compare the published
parameters with the estimates produced in this work --, and are sufficiently
dispersed throughout both galaxies. In Fig.~\ref{fig:ra-dec} we show their
spatial distribution.

Table~\ref{tab:literature} lists the nineteen articles that analyzed the same
$CT_1$ photometry used by \texttt{ASteCA} in the current study. Hereafter, we
will refer to this group as the ``literature''.
Metallicities, ages, reddenings, and distance moduli ($\mu_{\circ}$)
were estimated or assigned in the literature, except for the 36
clusters in~\cite{Piatti_2011b} which had only their ages estimated
via the $\delta T_1$ index \citep{Phelps_1994,Geisler_1997}.
In most of the literature metallicities and the distance moduli are fixed to
$[Fe/H]{=}-0.7$ dex, $[Fe/H]{=}-0.4$ dex, and $\mu_{\circ}{=}18.9$ mag and
$\mu_{\circ}{=}18.5$ mag, for the S/LMC respectively.
Ages reported in the literature were obtained by eye matching, either through
the standard isochrone technique or applying the $\delta T_1$ index method.
Reddenings were estimated in almost all cases interpolating the maps
of either~\cite{Burstein_1982},~\cite{Schlegel_1998}, or~\cite{Haschke_2011}.
\cite{Maia_2013} presented total mass estimations for their 29 star clusters
sample.

Our cluster sample was also partially studied via Johnson-Kron-Cousins
photometry, as listed in Table~\ref{tab:databases}. This group will be referred
as the ``databases'' (DBs), as a way to distinguish them from the
``literature''. In Sect.~\ref{sec:comp-pub-data} we compare the parameter
values obtained by \texttt{ASteCA}, to those given in both the literature and
the databases.

\begin{table} 
\centering
 \caption{Sources of the $CT_1$ data sets used in this work. $N$ refers to the
 number of clusters analyzed per article.}
\label{tab:literature}
 \begin{tabular}{l c c c}
\hline\hline
Article & $N$ & Galaxy & Telescope \\
\hline
\cite{Geisler_2003} & 8 & LMC & CTIO 0.9m \\ 
\cite{Piatti_2003a} & 5 & LMC & CTIO 0.9m \\ 
\cite{Piatti_2003b} & 6 & LMC & CTIO 0.9m \\ 
\cite{Piatti_2005} & 8 & SMC & CTIO 0.9m \\ 
\cite{Piatti_2007a} & 4 & SMC & CTIO 0.9m \\ 
\cite{Piatti_2007b} & 2 & SMC & Danish 1.54m \\ 
\cite{Piatti_2007c} & 2 & SMC & Danish 1.54m \\ 
\cite{Piatti_2008} & 6 & SMC & Danish 1.54m \\ 
\cite{Piatti_2009} & 5 & LMC & \vtop{\hbox{\strut CTIO 0.9m /}
                                     \hbox{\strut Danish 1.54m}} \\ 
\cite{Piatti_etal_2011a} & 3 & LMC & CTIO 0.9m \\ 
\cite{Piatti_etal_2011b} & 14 & SMC & CTIO 1.5m \\ 
\cite{Piatti_2011a} & 9 & SMC & Blanco 4m \\ 
\cite{Piatti_2011b} & 36 & LMC & Blanco 4m \\ 
\cite{Piatti_2011c} & 11 & SMC & Blanco 4m \\ 
\cite{Piatti_2012a} & 26 & LMC & Blanco 4m \\ 
\cite{Piatti_2012b} & 4 & SMC & Blanco 4m \\ 
\cite{Palma_2013} & 23 & LMC & Blanco 4m \\ 
\cite{Maia_2013} & 29 & SMC & Blanco 4m \\ 
\cite{Choudhury_2015} & 38 & LMC & Blanco 4m \\ 
\hline
 \end{tabular} 
\end{table}

\begin{table}
\centering
  \caption{Sources of Johnson-Kron-Cousins photometric data sets for some
  clusters in our sample. $N$ refers to the number of clusters in common with
  our sample per article.}
\label{tab:databases}
 \begin{tabular}{l c c c}
\hline\hline
Article & $N$ & Galaxy & Phot\\
\hline
\cite{Pietrzynski1999}, P99 & 7 & SMC & $BVI$ \\ 
\cite{Pietrzynski2000}, P00 & 25 & LMC & $BVI$ \\ 
\cite{Hunter_2003}, H03 & 62 & S/LMC & $UBVR$ \\ 
\cite{Rafelski_2005}, R05 & 24 & SMC & $UBVI$ \\ 
\cite{Chiosi_2006}, C06 & 16 & SMC & $VI$ \\ 
\cite{Glatt_2010}, G10 & 61 & S/LMC & $UBVI$ \\ 
\cite{Popescu_2012}, P12 & 48 & LMC & $UBVR$ \\ 
\hline
 \end{tabular} 
\end{table}


\section{Estimation of star cluster parameters}
\label{sec:fund-params}

Fundamental -- metallicity, age, distance modulus, reddening, mass -- and
structural -- center coordinates, radius, contamination index, approximate
number of members, membership probabilities, true cluster probability --
parameters, were obtained either automatically or semi-automatically with
\texttt{ASteCA}.
A detailed description of the functions built within this tool can be found in
Paper I, and in the code's online
documentation\footnote{\url{http://asteca.rtfd.org}}.
The resulting catalog can be accessed via
VizieR\footnote{\url{http://vizier.XXXX}}.
We have made available the Python codebase developed to analyze
the data obtained with \texttt{ASteCA}, and generate the figures in this
article.\footnote{\url{https://github.com/Gabriel-p/mc-catalog}}
Output images generated by \texttt{ASteCA} for each cluster, can be
accessed through a separate public code
repository.\footnote{\url{https://github.com/Gabriel-p/mc-catalog-figs}}


\subsection{Ranges for fitted fundamental parameters}
\label{ssec:param-ranges}

To processes a cluster, the user must provide \texttt{ASteCA} a
suitable range of accessible values for each fundamental parameter by
setting a minimum, a maximum, and a step.
As explained in Sect.~\ref{ssec:synth-match}, each combination of values
from the five fundamental parameters represents a unique synthetic CMD, or
model.
%
%
The larger the number of accessible parameter values, the larger the
amount of models the code will process to find the synthetic CMD which best
matches the observed cluster CMD.\@
Ranges and steps were selected to provide a balance between a large interval,
and a computationally manageable number of total models;
see Table~\ref{tab:ga-range}. Special care was taken to avoid defining ranges
that could bias the results towards a particular region of any fitted
parameter.\\

Unlike most previous works where the metallicity is a fixed value, we do not
make assumptions on the cluster's metal content. Our [Fe/H] interval covers
completely the usual metallicities reported for MCs' clusters.
The age range encompasses almost the entire allowed range of the CMD
service\footnote{\url{http://stev.oapd.inaf.it/cgi-bin/cmd}}
where the theoretical isochrones were obtained from (see
Sect.~\ref{ssec:synth-match}).

The maximum allowed value for the reddening of each cluster was determined
through the Magellanic Clouds Extinction Values (MCEV) reddening maps
\citep{Haschke_2011}\footnote{\url{
http://dc.zah.uni-heidelberg.de/mcextinct/q/cone/form}}, while
the minimum value is always zero.
%
We used TOPCAT\footnote{\url{http://www.star.bris.ac.uk/~mbt/topcat/}}
to query $E_{V-I}$ values from these maps, within a region as small as possible
around the position of each cluster.
For 85\% of our sample we found several regions with associated reddening
values, within a box of 0.5 deg centered on the cluster's position.
For the remaining clusters, larger boxes had to be used. The two most extreme
cases are NGC1997
($\alpha{=}5^h30^m34^s$, $\delta{=}-63^\circ12'12''$ [J2000.0]) and OHSC28
($\alpha{=}5^h55^m35^s$, $\delta{=}-62^\circ20'43''$ [J2000.0]) in the outers
of the LMC, where boxes of 4 deg and 6 deg respectively where needed to find a
region with assigned reddening values. In both cases, the reddenings given by
the~\cite{Schlafly_2011} map for their coordinates are up to two times smaller
than the ones found in the MCEV map.
We adopted the largest $E_{V-I}$ value of each region, MCEV$_{\max}$, as the
upper limit in the reddening range. Three steps are used to ensure that the
reddening range is partitioned similarly for all MCEV$_{\max}$ values: 0.01 for
MCEV$_{\max} {>}0.1$, 0.02 if $0.05{\leq}\mathrm{MCEV}_{\max}{\leq}0.1$, and
0.005 for MCEV$_{\max}{<}0.05$.
The $E_{V-I}$ extinction is converted to $E_{B-V}$
following~\cite{Tammann_2003}: $E_{V-I}{=}1.38\,E_{B-V}$. An extinction
law of $R_v{=}3.1$ is applied throughout the analysis.

Mean distance moduli for the S/LMC Clouds were taken
from~\cite{de_Grijs_2015} and~\cite{de_Grijs_2014}.
%
Line of sight (LOS) depths for the MCs (front to back, $\pm1\sigma$) have been
reported to span up to 20 kpc in their deepest
regions~\citep{Subramanian_2009,Nidever_2013,Scowcroft_2015}.
The 0.1 mag deviations allowed in this work give LOS depths of ${\sim}5.7$ kpc
and ${\sim}4.6$ kpc for the S/LMC\@. This covers more than half of the average
LOS depths found in~\cite{Subramanian_2009} for the SMC (bar: $4.9\pm1.2$ kpc,
disk: $4.23\pm1.48$ kpc), and the LMC (bar: $4.0\pm1.4$ kpc, disk: $3.44\pm1.16$
kpc).
Although this is not enough to cover the entire observed depth ranges, it gives
the distance moduli liberty to move around their mean values when all parameters
are adjusted.

The maximum total cluster mass was set to $10000\,M_{\odot}$.
%
To estimate this value, a first rough pass was performed with
\texttt{ASteCA} for all clusters in our dataset. Most of them were assigned
total masses below $5000\,M_{\odot}$, making the $10000\,M_{\odot}$ limit a
reasonable value.
This was true for all except 15 visibly massive clusters, for which the
maximum mass was increased to $30000\,M_{\odot}$. We will see
in Sect~\ref{sssec:integ_photom_masses} that the code is systematically
underestimating masses, due to the stellar crowding effect on our set of
Washington photometry.

Binary fraction was fixed to 0.5 -- considered a
reasonable estimate for clusters~\citep{von_Hippel_2005,Sollima_2010} -- to
avoid introducing an extra degree of complexity into the fitting process.
Secondary masses are randomly drawn from a uniform mass ratio distribution of
the form $0.7{\le}q{\le}1$, where $q{=}M_2/M_1$, and $M_1$, $M_2$ are the
primary and secondary masses. This range for the secondary masses was found for
the LMC cluster NGC1818 in~\cite{Elson_1998}, and represents a value
commonly used in analysis regarding the MCs~\citep[see][and references therein]
{Rubele_2011}.

\begin{table}
\centering
\caption{Fundamental parameters' ranges used by \texttt{ASteCA} on our set
of 239 clusters. The approximate number of values used for each parameter is
$N$. This gives a combined total of ${\sim}2.3 {\times}10^7$ possible models 
(or synthetic cluster CMDs), that could be theoretically matched to each studied
cluster in our sample.}
\label{tab:ga-range}
\begin{tabular}{lcccc}
\hline\hline
 Parameter & Min & Max & Step & $N$\\
\hline
[Fe/H] & $\sim$-2.2 & 0 & $\sim$0.1 & 23\\
$\log\mathrm{(age/yr)}$ & 6. & 10.1 & 0.05 & 82\\
$E_{B-V}$ & 0.0 & MCEV$_{\max}$ & ${\sim}$0.01 & $\sim$12\\
$\mu_{SMC}$ & 18.86 & 19.06 & 0.02 & 10\\
$\mu_{LMC}$ & 18.4 & 18.6 & 0.02 & 10\\
Mass ($M_{\odot}$) & 10 & [1, 3]${\times}10^{4}$ & 200 & [50, 150]\\
\hline
\end{tabular}
\end{table}


\subsection{Center and radius assignment}
\label{ssec:centre-radius}

\texttt{ASteCA} employs a two-dimensional Gaussian kernel density estimator 
(KDE) to determine the center of the cluster. A radial density profile (RDP) is
used to estimate the cluster's radius, as the point where the RDP reaches the
surrounding field's mean density.

The density of cluster members must make it stand out over the combination of
foreground/background stars, with no other over-density present in the observed
frame (this limitation is planned to be lifted in upcoming versions of
\texttt{ASteCA}). A portion of the surrounding field should be visible, and the
RDP must be reasonably smooth.
When these conditions are not met, the semi-automatic mode can be used. Here
center coordinates are either obtained based on an initial set of approximate
values, or manually fixed along with the radius value.

For ${\sim}66\%$ of our sample, center coordinates were obtained via a KDE
analysis, based on initial approximate values. Radii were calculated for
${\sim}60\%$ of the clusters, through an RDP analysis of their surrounding
fields.
The remaining clusters are those that are highly contaminated, contain very few
observed stars, and/or occupy most of the observed frames. This means that an
estimation of their centers and/or radii was not possible, and their values
were manually fixed.

The contamination index ($CI$) is a parameter related to the number of
foreground/background stars in the cluster region. See Paper
I, Sect. 2.3.2 for a complete mathematical definition of the index.
Values of $0,\,0.5$, and $\,{>}0.5$ mean respectively: the cluster is not
contaminated by field stars, an equal number of field stars and cluster
members are present in the cluster region, more field stars are
expected within the cluster region than cluster stars.
For reference, the average $CI$ for the set of clusters with manual radii
assignment is $CI{\simeq}0.9{\pm}0.2$, and $CI{\simeq}0.6\pm0.2$ for those
clusters whose radii were estimated in automatic mode.


\subsection{Field-star decontamination}
\label{ssec:dencontamination}

\begin{figure}
\centering
\includegraphics[width=\hsize]{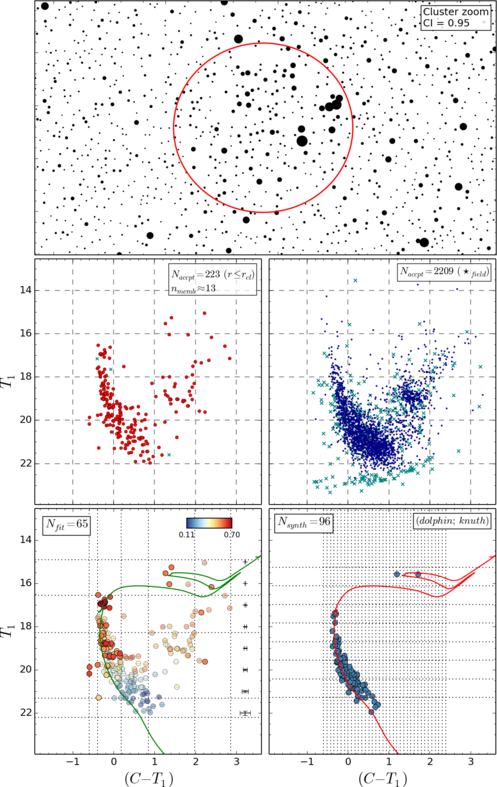}
\caption{\emph{Top}: SMC-L62 cluster and surrounding field stars
region. The adopted radius is marked with a red circle. The CI value in the top
right box is the ``contamination index'', see Sect.~\ref{ssec:centre-radius}.
\emph{Middle}: CMD of the cluster region for SMC-L62 (left),
where $n_{memb}$ is the approximate number of cluster stars. The CMD
of ten combined surrounding field regions, each with an area equal to
that of the cluster, is also shown (right).
In both panels $N_{accpt}$ is the number of stars that were not rejected due to
their large photometric errors. Rejected stars with large errors are shown as
pale green crosses.
\emph{Bottom}: cluster region after applying the DA (left); MPs vary
according to the colorbar at the top right. Dotted horizontal and vertical lines
show the binning used to reject low MP stars cell-by-cell, as obtained via the
Bayesian blocks method. Open circles (drawn semi-transparent) represent rejected
stars.
$N_{fit}$ is the number of stars kept unsubtracted by the cell-by-cell
rejection. The best fitted isochrone is overplotted with a green line. 
The respective generated synthetic CMD is shown in the right panel. $N_{synth}$
is the number of stars in the synthetic cluster CMD, and the dotted lines
represent the binning obtained using Knuth's rule, applied in the synthetic
cluster match process. The limiting magnitude of the synthetic CMD is
taken from the limiting magnitude of the observed cluster (roughly
$T_1{\approx}22$ mag, in this case).}
\label{fig:DA_BF}
\end{figure}

A decontamination algorithm (DA) was employed on the $CT_1$ CMD of each
processed cluster to remove field-star contamination.
The Bayesian DA presented in Paper I was improved for the present analysis; the
new DA works in two steps. First, the original Bayesian membership probability 
(MP) assignation is applied to the CMD of all stars within the cluster region 
(see Paper I for more details).
After that the CMD is binned into cells, and a cleaning algorithm is used
to remove stars of low MPs cell-by-cell as shown in Fig.~\ref{fig:DA_BF}.
By default \texttt{ASteCA} uses the Bayesian blocks method\footnote{\url
{http://www.astroml.org/examples/algorithms/plot_bayesian_blocks.html}}
introduced in~\cite{Scargle_2013}, via the implementation of the astroML
package~\citep{Vanderplas_2012}. While Bayesian blocks binning is the
default setting in \texttt{ASteCA}, several others techniques for CMD star
removal are available, as well as five more binning methods.
This second step is similar to the DA developed in~\citet[][B07]{Bonatto_2007},
which uses a simpler rectangular grid. The main difference, aside from the
binning method employed, is that \texttt{ASteCA} removes stars based on their
MPs, not randomly as done in B07.

Approximately 70\% of our sample was processed with these default settings.
The remaining ${\sim}$30\%, those with a low number of cluster stars
or heavily contaminated, were processed with modified settings to allow a proper
field-star decontamination.
Changes introduced were, for instance, a different binning
method~\citep[often a rectangular grid using Scott's rule,][]{Scott_1979},
or skipping the Bayesian MP assignation and only performing a density based
cell-by-cell field star removal. In this latter case, the DA works very
similarly to the B07 algorithm.

An appropriate field-star decontamination is of the utmost importance, since the
cleaned cluster CMDs will be used to estimate the cluster fundamental parameters.


\subsection{Synthetic CMD matching}
\label{ssec:synth-match}

\texttt{ASteCA} derives clusters' fundamental parameters matching their observed
CMDs with synthetic CMDs. In this work these are built using PARSEC v1.1
theoretical isochrones~\citep[][B12]{Bressan_2012},
and a log-normal initial mass function~\citep[IMF,][]{Chabrier_2001}.
For a given age, metallicity, total mass, and binary fraction, a
synthetic CMD is built from a stochastically sampled IMF down to the
faintest portion of the theoretical isochrone (defined by the metallicity and
age values), shifted by a reddening and distance modulus. For a detailed
description on the generation of synthetic CMDs from theoretical
isochrones by \texttt{ASteCA}, see: Paper I, Sect. 2.9.1. All the evolutionary
tracks from the CMD service\footnote{\url{http://stev.oapd.inaf.it/cgi-bin/cmd}}
are currently supported, as well as three other IMFs.

The Poisson likelihood rate~\citep[PLR,][]{Dolphin_2002} is employed
to asses the match between the cluster's CMD and a synthetic
CMD, out of the ${\sim}2{\times}10^7$ possible solutions (as shown in
Table~\ref{tab:ga-range}).
The PLR statistic requires binning the cluster's CMD, and the synthetic CMDs
generated according to the fundamental parameter ranges defined in
Sect.\ref{ssec:param-ranges}.
We use Knuth's rule~\citep[][also implemented via the astroML package]
{Knuth_2006} as the default binning method (see bottom right plot in
Fig.~\ref{fig:DA_BF}).
The inverted logarithmic form of the PLR can be written as

\begin{equation}
LPLR  = -2 \ln PLR = 2 \sum_i m_i - n_i + n_i \ln \frac{n_i}{m_i},
\label{eq:likelihood}
\end{equation}

\noindent where $m_i$ and $n_i$ are the number of stars in the $i$th cell
(two-dimensional bin) of the synthetic and the observed cluster's CMD,
respectively. If for any given cell we have $n_i\neq0$ and $m_i=0$, a
very small number is used instead ($m_i=1{\times}10^{-10}$) to avoid a
mathematical inconsistency with the factor $\ln m_i$.
Although \texttt{ASteCA} does not currently provide a goodness-of-fit
estimator, uncertainties associated to the fitted fundamental parameters
can be though of as a coarse measure of the fit's robustness. This parameter is
to be added in upcoming versions of the code.

In Paper I the total mass parameter could not be estimated, due to the
likelihood statistic used (Paper I, Eq. 11). The LPR defined in
Eq.~\ref{eq:likelihood} allows us to also consider the mass as a free parameter
in the search for the best synthetic CMD.\@
The total mass is estimated simultaneously along with the remaining
fundamental parameters of a cluster, no extra process is employed (for example,
a mass-luminosity relation).
Following the validation performed in Paper I for the metallicity, age,
reddening, and distance, we present in Appendix~\ref{apdx:mass_valid} a
similar study for the total mass. We demonstrate that the masses recovered by
\texttt{ASteCA} for nearly 800 MASSCLEAN synthetic clusters -- with
masses ranging from 500 $M_{\odot}$ to $2.5\times10^5\,M_{\odot}$ -- are in
excellent agreement with the masses used to generate them.

The determination of any fundamental parameter depends exclusively on
the distribution and number of observed stars in the cluster's CMD.\@
Since we deal with five free fundamental parameters, a 5-dimensional surface of
solutions is built from all the possible synthetic CMDs.
\texttt{ASteCA} applies a genetic algorithm (GA) on this surface to derive the
cluster's fundamental parameters.
%
%
After the GA returns the optimal fundamental parameter values, uncertainties are
estimated via a standard bootstrap technique. This process takes a significant
amount of time to complete, since it involves running the GA several more times
on a randomly generated cluster with replacement. Generating a new
cluster ``with replacement'', means randomly picking stars one by one from the
original cluster, where any star can be selected more than once. The process
stops when the same number of stars as those present in the original cluster
have been picked.
We run the bootstrap ten times per cluster, as it would be prohibitively
costly timewise to run it -- as would be ideal -- hundreds or even thousands of
times.
On average, the CMD of each cluster in our dataset was compared t
${\sim}1{\times}10^6$ synthetic CMDs, once the full process was completed.


\section{Errors in fitted parameters}
\label{sec:errors-fit}

It is well known that a parameter given with no error estimation is
meaningless from a physical standpoint~\citep{Dolphin_2002,Andrae_2010}.
%
Despite this, a detailed error treatment is often ignored in
articles that deal with star clusters analysis~\citep{Paunzen_2006}.

As explained in Sect.~\ref{ssec:synth-match}, \texttt{ASteCA} employs a
bootstrap method to assign standard deviations for each fitted parameter.
The code simultaneously fits five free parameters within a wide range of allowed
values, using only a 2-dimensional space of observed data, i.e., the $T_1$
versus $ (C-T_1)$ CMD;\@ meaning uncertainties will be somewhat large.
We plan to upgrade the code to eventually allow more than just two observed
magnitudes, extending the 2-dimensional CMD analysis to an N-dimensional one.
It is worth noting that, unlike manually set errors, these are
statistically valid uncertainty estimates. This is an important point to make
given that the usual by-eye isochrone fitting method not only disregards known
correlations among all clusters parameters, it is also fundamentally incapable
of producing a valid error analysis~\citep{Naylor_2006}. Any uncertainty
estimate produced by-eye serves only as a mere approximation, which will often
be biased towards smaller figures. The average logarithmic age error given in
the literature, for example, is 0.16 dex, in contrast with the almost twice as
large average value estimated by \texttt{ASteCA} (see below).
In Fig.~\ref{fig:errors} we show the distribution of standard deviations
of the five fundamental parameters fitted by the code, for the entire sample.

\begin{figure}[!ht]
\centering
\includegraphics[width=\hsize]{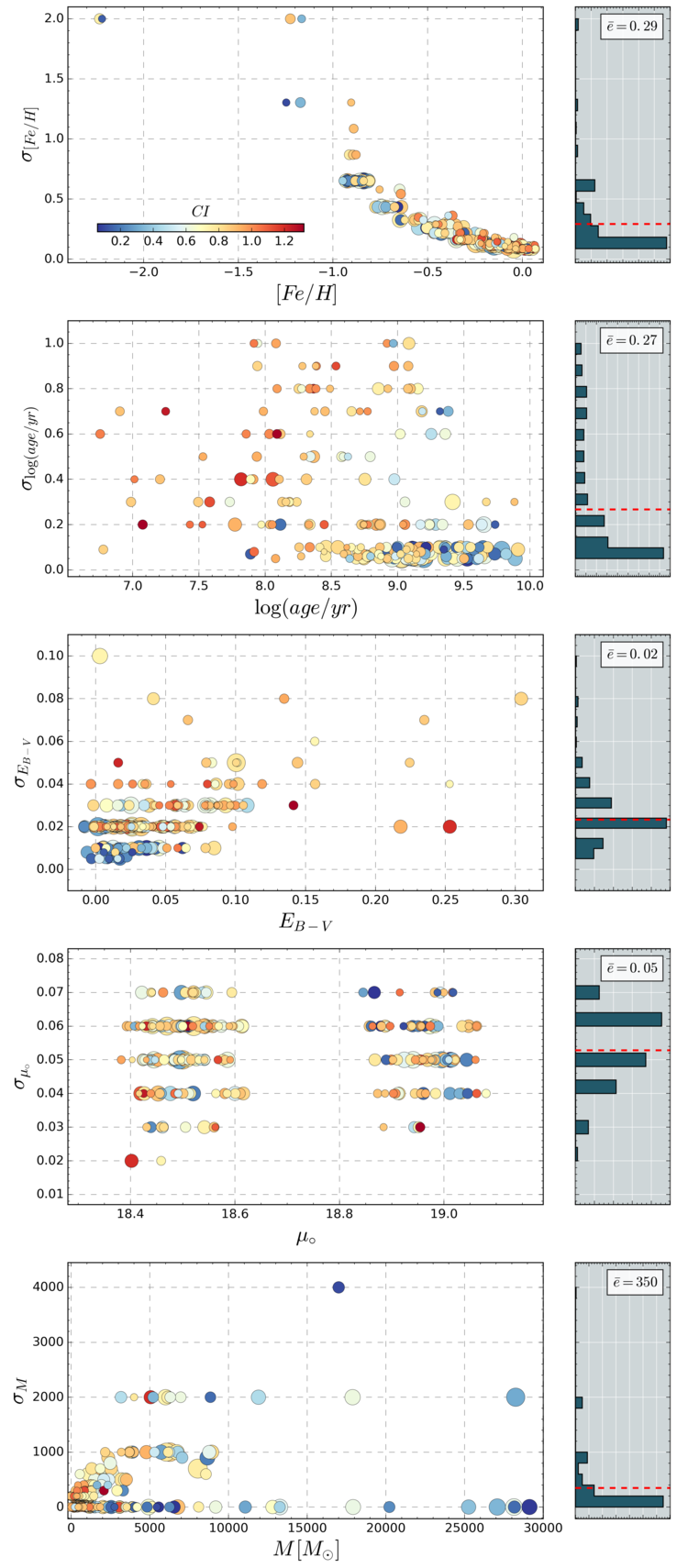}
\caption{\emph{Left}: Distribution of errors versus the five parameters fitted
by \texttt{ASteCA}.\@ Colors are associated to the $CI$ (see bar in top plot),
sizes are proportional to the actual cluster sizes. A small random scatter in
the x-axis is added for clarity.
\emph{Right}: Error histogram. The mean error value for each parameter is shown
in the top right corner, and drawn in the plot with a dashed red line.}
\label{fig:errors}
\end{figure}

The apparent dependence of the metallicity error with decreasing [Fe/H] values
arises from the fact that \texttt{ASteCA} uses $z$ values, where
$\mathrm{[Fe/H]}{=}\log(z/z_{\odot})$; with $z_{\odot}{=}0.0152$. We use $z$
instead of [Fe/H] because: a- this is the default form in which the evolutionary
tracks are generated by the CMD service, and b- it is easier to work with the
code when all parameters are strictly positive.
The error in $z$ is $\sigma_z{\approx}0.003$ for over 75\% of the clusters
analyzed. This means that, when converting to $e_{[Fe/H]}$ using the relation

\begin{equation}
\sigma_{[Fe/H]} = \sigma_z/[z\times\ln(10)],
\end{equation}

\noindent the $z$ in the denominator makes $\sigma_{[Fe/H]}$ grow as it
decreases, while $\sigma_z$ remains approximately constant.
For very small $z$ values (e.g.: $z{=}0.0001$), the logarithmic errors can
surpass $\sigma_{[Fe/H]}{=}2$ dex. In those cases the error is trimmed to 2 dex,
enough to cover the entire metallicity range.

There are no visible error trends for any of the fitted parameters, a desirable
feature for any statistical method. If a parameter's uncertainty varied 
(increase/decrease) with it, it would indicate \texttt{ASteCA} was
introducing biases in the solutions.

Histograms to the right of Fig.~\ref{fig:errors} show the distribution
of errors, and their arithmetic means as a dashed red line.
For $\sim$34\% and ${\sim}69\%$ of the S/LMC clusters we have
$\sigma_{[Fe/H]}{<}0.2$ dex.
Approximately 53\% of the combined S/LMC sample show
$\sigma_{\log(age/yr)}{<}0.1$ dex. Error estimates for the remaining parameters
are all within acceptable ranges. The uncertainties tend to increase for
clusters of lower mass, due to the inherent stochasticity of the fitting
process.\\

Errors could be lowered applying different techniques: increase the
number of models evaluated in the GA, reduce the value of the steps in the
parameters' range, or increase the number of bootstrap runs.
These methods, particularly the latter, will extend the time needed to process
each cluster. Limited computational time available requires a balance between
the maximum processing power allocated to the calculations, and the aimed
precision. Error values presented here should then be considered a conservative
upper limit.


\section{Comparision with published fundamental parameters}
\label{sec:comp-pub-data}

We compare in Sect.~\ref{ssec:lit-values} the parameter values obtained by
\texttt{ASteCA}, with those taken from the papers that used the same
$CT_1$ data sets (see Table~\ref{tab:literature}), referred as the
``literature''.
The parameters age, reddening, and mass are also analyzed in
Sect~\ref{ssec:db-values} for a subset of 142 star clusters that could be
cross-matched with $UBVRI$ photometry results (see Table~\ref{tab:databases}),
referred as the ``databases''.


\subsection{Literature values}
\label{ssec:lit-values}

\begin{figure*}
\includegraphics[width=2.\columnwidth]{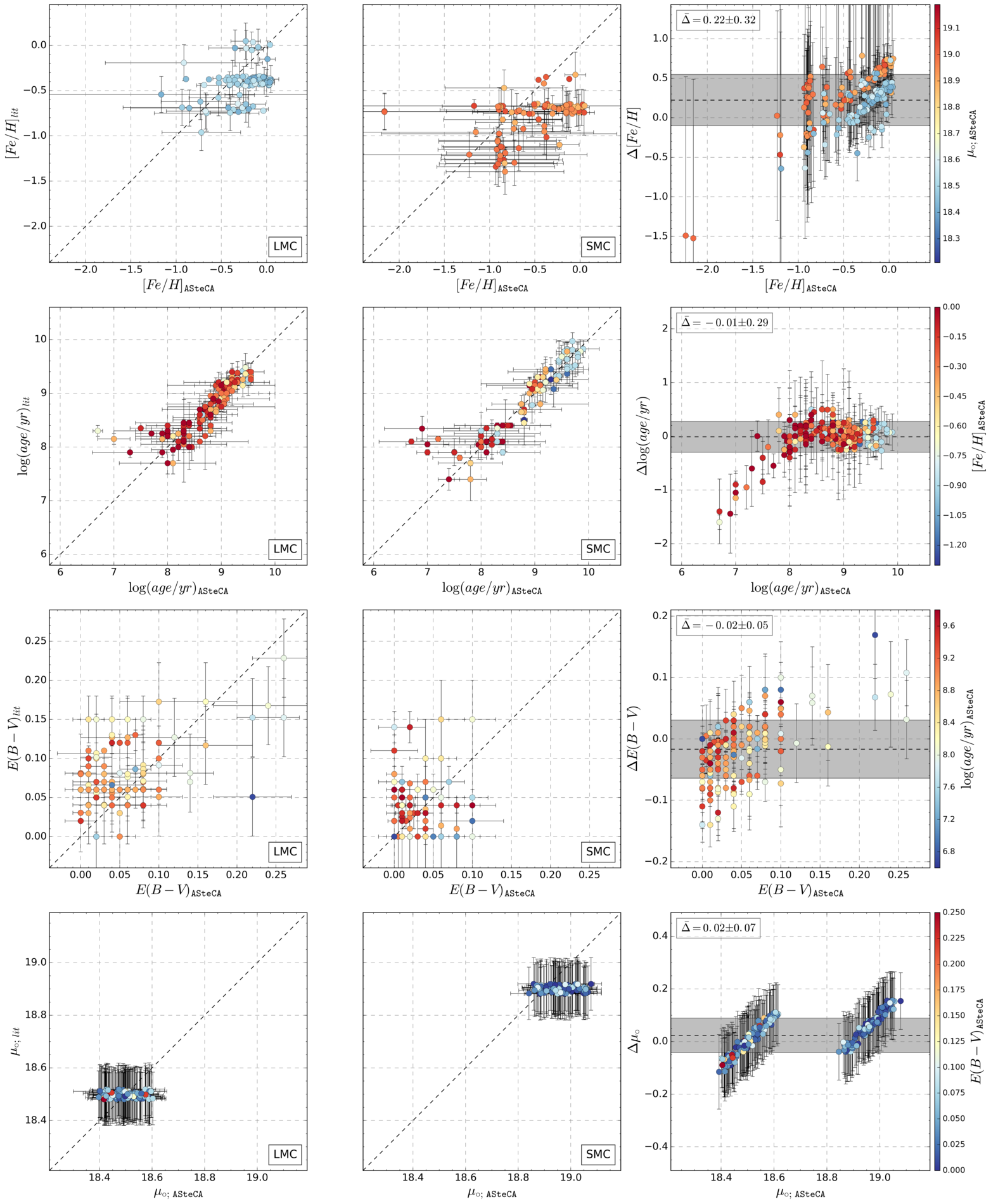}
\caption{\emph{Left column}: parameters comparison for the LMC.\@
\emph{Center column}: idem for the SMC.\@
\emph{Right column}: BA plot with differences in the sense $\Delta$= (\texttt
{ASteCA} minus literature), for the combined S/LMC sample.
A small random scatter is added to both axes for the metallicity and distance
modulus plots. Mean and standard deviation of the differences,
$\overline{\Delta}$, are shown as a dashed line and a gray band, respectively;
their values are displayed in the top left of the plot.
Colors follow the bar at the right of the figure, for each row.~\cite
{Piatti_2011b} clusters with age information only are plotted
with $E_{B-V}{=}0$.}
\label{fig:as_vs_lit}
\end{figure*}

\texttt{ASteCA} versus literature values for the metallicity, age, reddening,
and distance modulus, are presented in Fig.~\ref{fig:as_vs_lit}. Left and
central panels show the 1:1 identity line for both galaxies.
Right panel diagram shows a Bland-Altman (BA) plot for our combined sample
of clusters, with the variation in the x-axis proposed by
Krouwer~\citep{Bland_1986,Krouwer_2008}. The BA is also called a
``difference'' or ``Tukey Mean-Difference'' plot. In the original BA plot the
default x-axis displays the mean values between the two methods being compared.
The Krouwer variation changes the means for the values of one of those methods,
called the ``reference''. In our case, the reference is \texttt{ASteCA}
so we use its reported values in the x-axis.\\
%

\texttt{ASteCA} abundance estimates are slightly larger than those in
the literature, seen as offsets in the upper panels of Fig.~\ref{fig:as_vs_lit}.
On average, the offset is $\sim$0.27 dex and $\sim$0.18 dex for the S/LMC.\@
This effect can be explained by two different processes, in light of our
knowledge that the \texttt{ASteCA}'s best fit CMD matching introduces no
biases into the solutions.
First, \texttt{ASteCA}'s $z$ values are converted to the logarithmic
form [Fe/H] using a solar metallicity of
$z_{\odot}{=}0.0152$~\citep{Bressan_2012}.
Literature values, on the other hand, are converted using a solar
metal content of $z_{\odot}{=}0.019$~\citep{Marigo_2008} (usually
rounded to 0.02). This means that \texttt{ASteCA} will give [Fe/H] values
larger by ${\sim}0.1$ dex, for any fitted isochrone of equivalent $z$ being
compared, following

\begin{equation}
\begin{split}
\Delta\mathrm{[Fe/H]} & = \mathrm{[Fe/H]}_{\mathtt{ASteCA}} -
\mathrm{[Fe/H]}_{literature} \\
& = \log(z/0.0152) - \log(z/0.019) = \log(0.019/0.0152) \\
& \approx 0.0969 \;\mathrm{dex}
\end{split}
\label{eq:delta_feh}
\end{equation}

\noindent Second, the MC's star clusters are often studied
using fixed metallicities. For example, [Fe/H] values assumed in the 19
literature articles, are exactly -0.7 dex and -0.4 dex for $\sim$60\% and
$\sim$75\% of S/LMC clusters. The by-eye fit is biased towards the
assignment of these particular abundances, due to the effect
of ``confirmation bias''. This effect present in the published
literature has been studied recently by~\cite{de_Grijs_2014}, in relation to
distance measurement reported for the LMC.\@
Alternatively, the code is left free to fit metal contents in the
entire range given. The [Fe/H] values estimated by \texttt{ASteCA} are
distributed mainly between -1 dex and 0 dex, but clustered closer to solar
metal content than to lower metallicities. This fact, combined with the above
mentioned effect of mostly fixed [Fe/H] values used in the literature, makes the
averaged difference in assigned metal content be slightly positive.
It also is responsible for the apparent linear growth of $\Delta$[Fe/H] as
$\mathrm{[Fe/H]}_{\mathtt{ASteCA}}$ increases, seen in Fig.~\ref{fig:as_vs_lit}.
Researchers tend to fit isochrones adjusting it to the lower envelope of a
cluster's sequence. This is done to avoid the influence of the binary
sequence (located above and to the right of the single stars sequence in a CMD)
in the fit, but it can also contribute to the selection of
isochrones of smaller metallicity. The reason is that increasing an
isochrone's metal content moves it towards redder (greater color) values in a
CMD, see for example~\cite{Bressan_2012}, Fig 15.
These combined effects explain the offset found for the abundances. It is
important to notice that neither effect is intrinsic to the best likelihood
matching method used by the code.

The general dispersion between literature and \texttt{ASteCA} values can be
quantized by the standard deviation of the $\Delta$ differences in the BA plot.
This value is $\sim$0.32 dex for the metallicity, in close agreement with the
mean internal [Fe/H] uncertainty found in Sect.~\ref{sec:errors-fit}.
\texttt{ASteCA}'s mean metallicities for the MCs are
$\mathrm{[Fe/H]}_{SMC}{\simeq-}0.52{\pm}0.44$ dex, and
$\mathrm{[Fe/H]}_{LMC}{\simeq-}0.26{\pm}0.24$ dex. These averages are similar,
within their uncertainties, to those obtained using literature values:
$\mathrm{[Fe/H]}_{SMC}{\simeq-}0.78{\pm}0.23$ dex, and
$\mathrm{[Fe/H]}_{LMC}{\simeq-}0.42{\pm}0.16$ dex\\

Ages show an overall good agreement, with larger differences seen for a handful
of young clusters. Ten clusters with either a shallow photometry, a low number
of cluster stars, or heavily contaminated, present age differences larger than
$\Delta \log(age/yr){>}0.5$ dex. These are referred to as ``outliers'', and
discussed in more depth in Appendix~\ref{apdx:outliers}.
%
%
The mean $\Delta \log(age/yr)$ value for all S/LMC clusters, -0.01 dex as shown
in the age BA plot of Fig.~\ref{fig:as_vs_lit}, points to an excellent agreement
in $\log(age/yr)$. Excluding outliers, this mean increases to $\sim$0.04 dex.

Similarly to what was found for the metallicity, the $\sim$0.3 dex dispersion
is almost exactly the internal uncertainty found for errors assigned by the
code.\\

The reddening distribution presents maximum $E_{B-V}$ values of $\sim$0.15
mag and $\sim$0.3 mag for the S/LMC, respectively.
The $\Delta$ differences are well balanced with a standard deviation of 0.05
mag, slightly larger than the 0.02 mag average uncertainty found for internal
errors. We estimate average $E_{B-V}$ values for the S/LMC of $0.03{\pm}0.03$
mag and $0.05{\pm}0.05$ mag.
These estimates are approximately a third of those used for example in
the~\cite{Hunter_2003} study, because our sample does no contain clusters in the
regions of the MCs most affected by dust.\\

Distance moduli ($\mu_{\circ}$) found by \texttt{ASteCA} show a clear
displacement from literature values. This is expected, as the clusters' distance
in the literature articles is always assumed equivalent to the
distance to the center of the corresponding galaxy.
The distribution of $\mu_{\circ}$ values found by the code covers the entire
range allowed in Sect.~\ref{ssec:param-ranges}.
%
It is worth noting that this variation appears to have no substantial effect on
any of the remaining parameters, something that could in principle be expected
due to the known correlations between them (see Paper I, Sect. 3.1.4).
This reinforces the idea that using a fixed value for the distance modulus, as
done in the literature, is a valid way of reducing the number of free variables
at no apparent extra cost.\\

\begin{figure}
\centering
\includegraphics[width=\hsize]{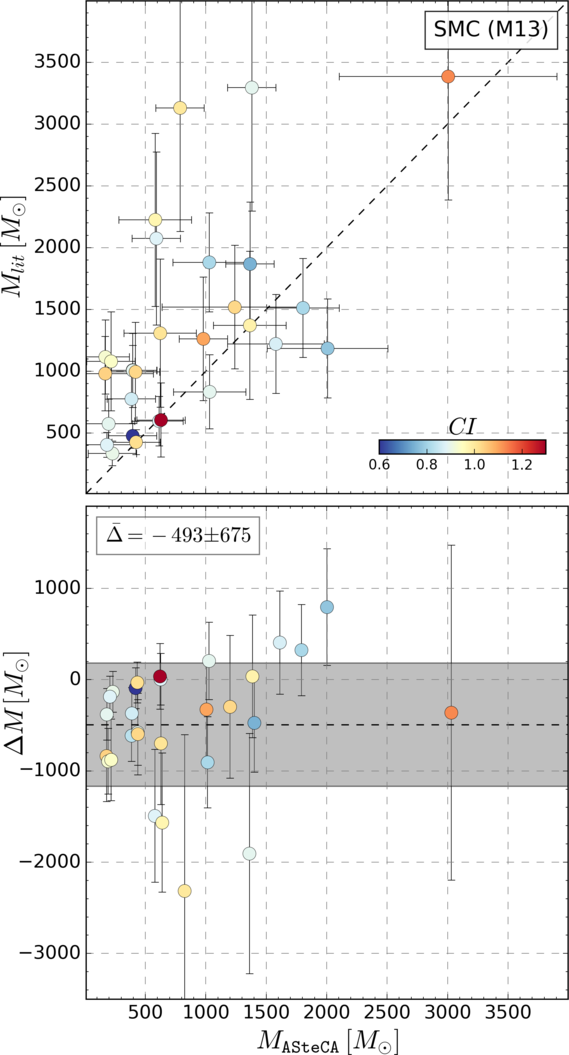}
\caption{\emph{Top}: Mass comparison for \texttt{ASteCA} versus literature
values.
\emph{Bottom}: BA plot with mean and standard deviation of the differences
shown as a dashed horizontal line, and a gray band respectively. The mean and
standard deviation values are displayed in the top left corner.}
\label{fig:as_vs_lit_mass}
\end{figure}

Masses are only assigned in~\citet[][M13]{Maia_2013} for its sample of 29
SMC clusters. The comparison with \texttt{ASteCA} is presented in
Fig.~\ref{fig:as_vs_lit_mass}, where the top panel shows a trend for
\texttt{ASteCA} masses to be smaller than those from M13. This is particularly
true for the clusters with the largest mass estimates -- and the largest
assigned errors -- in M13: H86-97 (3300$\pm$1300$M_{\odot}$), and H86-87
(3100$\pm$1700$M_{\odot}$). The BA plot (bottom) shows that, on average, M13
masses are ${\sim}500{\pm}700\,M_{\odot}$ larger.
To explain these differences, we need to compare the way total cluster masses
are obtained by M13 and \texttt{ASteCA}.

\texttt{ASteCA} estimates masses matching synthetic CMDs to that of
the observed cluster region.
The radius that delimits this region is formally equivalent to the
tidal radius defined in~\cite[][see Sect.~\ref{ssec:centre-radius}]{King_1962},
but it is in practice smaller due to observational effects (i.e, Poisson
errors in star counts, photometric incompleteness, field stars contamination).
A fraction of cluster members is thus left out of the analysis when this radius
is used. We can prove using King's derivation of the number of
members for a cluster~\citep[Eq. 18,][]{King_1962}, that this fraction is small
except for highly underestimated radii ($r{<}0.5r_t$) and very low concentration
clusters ($r_t/r_c{<}2$).
Massive clusters are largely unaffected by this process, while uncertainties for
low-mass clusters' masses are large enough that this effect can be safely
ignored. To err on the side of caution, \texttt{ASteCA} masses should be
considered lower estimates of the true present-day cluster masses.

In M13 cluster masses were determined using two methods, based on estimating
a mass function via the $T_1$ luminosity function (LF).
In both cases a field star cleaning process was applied. The first one employs
a CMD decontamination procedure~\citep{Maia_2010}, and the
second one cleans the cluster region's LF by subtracting it a field star LF.\@
Results obtained with these two methods are averaged to generate the final
mass values.
Although any reasonable field star cleaning algorithm should remove many or most
of the foreground/background stars in a cluster region, some field stars are
bound to remain. For heavily contaminated clusters this effect will be
determinant in shaping the ``cleaned'' CMD sequence, as cluster stars will be
very difficult to disentangle from contaminating field stars.
Clusters in M13 are indeed heavily affected by field star contamination,
as seen in Fig.~\ref{fig:as_vs_lit_mass} where colors follow their $CI$s.
The minimum value is $CI{\simeq}0.55$, meaning all cluster regions are expected
to contain, on average, more field stars than cluster stars.
A large number of contaminating field stars not only makes the job much harder
for the DA, it also implies that the LF will most likely be overestimated. This
leads -- in the case of M13 --  to an overestimation of the total mass.
In contrast, \texttt{ASteCA} assigns masses taking their values directly from
the best match synthetic CMD.\@ Field star contamination will thus have a lower
influence on the code's mass estimate, limited just to how effective the DA is
in cleaning the cluster region.
We find in Appendix~\ref{apdx:mass_valid} that the code will slightly
underestimate masses by approximately 200 $M_{\odot}$, for low mass
clusters. This is unrelated to the fraction of members lost due
to the employed radius (mentioned previously), since we use in this analysis
the full tidal radius to delimit the MASSCLEAN synthetic clusters region.
These combined effects explain the ${\sim}500\,M_{\odot}$ offset found
for mass values, seen in Fig.~\ref{fig:as_vs_lit_mass}.

\begin{figure}
\centering
\includegraphics[width=\hsize]{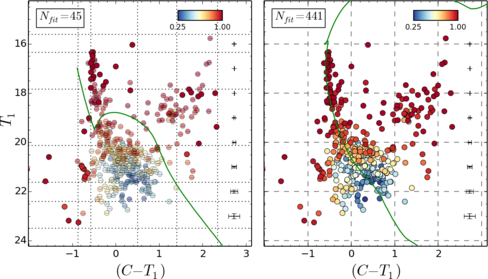}
\caption{\emph{Left}: Best fit isochrone for B48 found by \texttt{ASteCA} when
a cell-by-cell removal is applied, following the Bayesian MP assignation 
(removed stars are drawn semi-transparent). Estimated age and
total mass are $\log(age/yr){=}6.2{\pm}0.6$, and $M{=}400{\pm}200\,M_{\odot}$.
\emph{Right}: Best fit isochrone found when no removal of stars is performed,
and the full cluster region is used in the search for the best synthetic
cluster match. Estimated age and total mass are now
$\log(age/yr){=}7.5{\pm}0.3$, and $M{=}3000{\pm}900\,M_{\odot}$.}
\label{fig:B48_DA}
\end{figure}

The case of B48 is worth mentioning, as it is the cluster with the largest total
mass given in M13 (3400$\pm$1600$M_{\odot}$).
After removing possible field stars -- see Sect.~\ref{ssec:dencontamination} --
low mass stars disappear, and B48 is left only with its upper sequence 
($T_1<18.4$ mag). This happens both in M13, see Fig. C8, and the
\texttt{ASteCA} analysis, see left CMD in Fig.~\ref{fig:B48_DA}.
The likelihood (Eq.~\ref{eq:likelihood}) sees then no statistical benefit in
matching the cluster's CMD with a synthetic CMD of similar age and mass,
which will contain a large number of low mass stars.
This leads the GA to select synthetic CMDs of considerably younger ages
($\log(age/yr){<}7.0$ dex) than that assigned in M13
($\log(age/yr){=}7.9\pm0.05$ dex), and with much lower mass estimates (see
caption of Fig.~\ref{fig:B48_DA}).
%
%
Age and mass values similar to those from M13 could be found by the code, only
if the DA was applied with no cell-by-cell removal of low MP stars as shown in
the right CMD of Fig.~\ref{fig:B48_DA}. This means that all field stars within
the cluster region are used in the synthetic CMD matching process, which
inevitably questions the reliability of the total mass estimate.
Dealing with this statistical effect is not straightforward and will probably
require an extra layer of modeling added to the synthetic CMD generation
algorithm.
As discussed in Appendix~\ref{apdx:outliers} this effect also plays an
important role in the significant age differences found between \texttt{ASteCA}
and the literature, for a handful of outliers.

\subsection{Databases values}
\label{ssec:db-values}

\begin{figure*}
\includegraphics[width=2.\columnwidth]{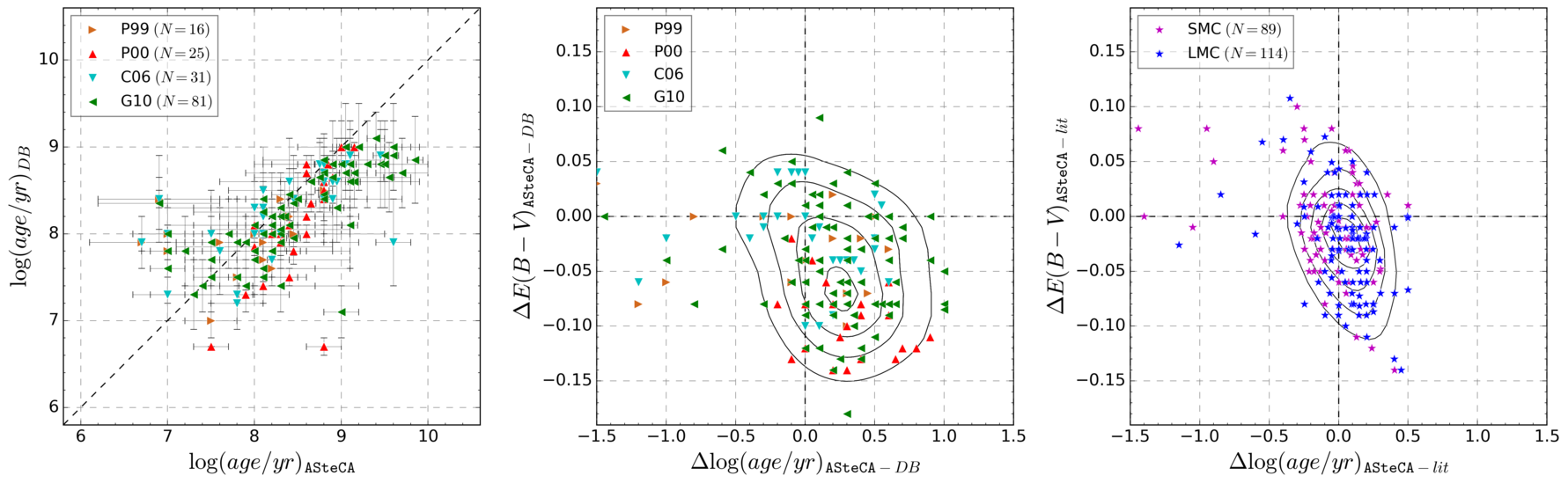}
\caption{\emph{Left}: age comparison for DBs that used the isochrone fit
method versus \texttt{ASteCA}, where $N$ is the number of clusters
cross-matched in each DB.\@
\emph{Center}: ``delta'' plot (\texttt{ASteCA} minus DB) showing differences
between reddening and age.\@ Curves represent regions of iso-densities after
fitting a 2-dimensional Gaussian Kernel.
\emph{Right}: same as previous plot, for \texttt{ASteCA} minus literature values
for both Clouds.}
\label{fig:cross_match_if}
\end{figure*}

We compare our results with those from seven articles -- the ``databases'' or
DBs -- where a different photometric system was used; see Table~\ref
{tab:databases}.
These DBs are separated into two groups: those where the standard by-eye
isochrone fitting method was applied -- P99, P00, C06, and G10 -- and those
where integrated photometry was employed -- H03, R05, and P12.
A total of 142 clusters from our sample could be cross-matched.
Where names were not available to perform the cross-match -- P00, P99, and R05
-- we employed a 20 arcsec finding radius, centered on the equatorial
coordinates of the clusters.\\

%
%
\texttt{ASteCA} versus the four isochrone-fit DBs ages, are shown in
Fig.~\ref{fig:cross_match_if}, left and center plot.
P99 and P00 analyze SMC and LMC clusters respectively,
using~\cite{Bertelli_1994} isochrones and fixed S/LMC metallicities
$z{=}0.004, 0.008$. While P99 derives individual reddening estimates based on
red clump stars, P00 uses reddening values determined for 84 lines-of-sight
in the~\cite{Udalski_1999} LMC Cepheids study.
%
Both studies attempt to eliminate field star contamination following the
statistical procedure presented in~\cite{Mateo_1986}.
These DBs employed distance moduli of 18.65 mag and 18.24 mag for the S/LMC,
approximately ${\sim}0.25$ mag smaller than the canonical distances assumed for
each Cloud. This has a direct impact on their obtained ages.
%
In~\cite{de_Grijs_2006} the authors estimate that had P00 used a value
of $\mu_0{=}18.5$ mag instead, their ages would be ${\sim}0.2{-}0.4$
dex younger; the same reasoning can be applied to the P99 age estimates.
A similar conclusion is reached by~\cite{Baumgardt_2013}.
Notice that in the latter the authors correct the age bias that
arises in P00 due to the small distance modulus used, increasing P00 age
estimates by 0.2 dex. This is incorrect, ages should have been decreased by that
amount. 
P99 and P00 logarithmic ages are displaced on average from \texttt{ASteCA}
values (in the sense \texttt{ASteCA} minus DB) by $-0.13{\pm}0.6$ dex and $0.37
{\pm}0.5$ dex respectively, as seen in Fig.~\ref{fig:cross_match_if} left plot.
In the case of P99, the distance modulus correction would bring the age values
to an overall agreement with \texttt{ASteCA}, although with a large scatter
around the identity line.
P00 age values on the other hand, would end up ${\sim}0.7$ dex below the code's
age estimates after such a correction. Such a large deviation is most likely due
to the overestimated reddening values used by P00, as will be shown below.

C06 studied 311 SMC clusters via isochrone fitting applying two methods: visual
inspection and a Monte Carlo based $\chi^2$ minimization. The authors also
employ a decontamination algorithm, making this the article that more closely
resembles our work.
Distance modulus is fixed to $\mu_0{=}18.9$ mag, reddening and
metallicity values of $E_{B-V}{=}0.08$ mag and $z{=}0.008$ are used, adjusted
when necessary to improve the fit.
It is worth noting that the [Fe/H]${=}-0.4$ dex abundance employed in
C06 is closer to the [Fe/H]${=}-0.52{\pm}0.44$ dex \texttt{ASteCA} average
for the SMC, than the canonical [Fe/H]${=}-0.7$ dex value used in most works.
Out of the seven DBs, C06 is the one that best matches \texttt{ASteCA}'s
$\log(age/yr)$ values, with a mean deviation from the identity line of
$0.02\pm0.58$ dex.

%

G10 analyzed over 1500 clusters with ages ${<}1$ Gyr in both Clouds via
by-eye isochrone fitting. They assumed distance moduli of (18.9, 18.5) mag, and
metallicities of (0.004, 0.008), for the S/LMC respectively. Reddening was
adjusted also by-eye on a case-by-case basis.
This DB presents a systematic bias where smaller logarithmic ages are
assigned compared to our values, with an approximate deviation of $\Delta
\log(age/yr){\simeq}0.23\pm0.46$ (\texttt{ASteCA} minus G10). This is consistent
with the results found in~\cite{Choudhury_2015} (see Fig. 5), and later
confirmed in~\cite{Piatti_2015a,Piatti_2015b}.
G10 does not apply any decontamination method, instead they plot a sample of
surrounding field stars over the cluster region. The lack of a proper
statistical removal of contaminating foreground/background stars can
skew the isochrone fit.

As seen in Fig.~\ref{fig:cross_match_if} (center plot) these four DBs
present a clear age-extinction bias compared to \texttt{ASteCA} values, with the
maximum density located around $\Delta E_{B-V}{\simeq-}0.07$ mag and
$\Delta\log(age/yr){\simeq}0.3$ dex.
This degeneracy was found in Paper I (Table 3) to have the largest
correlation value, meaning it is the process most likely to affect isochrone fit
studies.
The trend is most obvious for P00 where a large average reddening of
$E_{B-V}{\simeq}0.14$ mag~\citep{de_Grijs_2006} was employed, compared to the
mean value found by \texttt{ASteCA}.
Right of Fig.~\ref{fig:cross_match_if} we see the same plot,
generated subtracting literature from \texttt{ASteCA} age estimations.
The afore mentioned bias is basically non-existent here, pointing to a
consistent assignation of reddening and ages by the code.

In Appendix~\ref{apdx:databases} we show CMDs for the 153 cross-matched
clusters across these four DBs.
Clusters that present the largest age discrepancies between
\texttt{ASteCA} and the DBs, are those where the same effect mentioned in
Appendix~\ref{apdx:outliers} takes place. A good example of this is SMC-L39, as
seen in Figs.~\ref{fig:DBs_C06_3} and~\ref{fig:DBs_G10_8} for C06 and G10
respectively.\\

%

\begin{figure}
\centering
\includegraphics[width=\hsize]{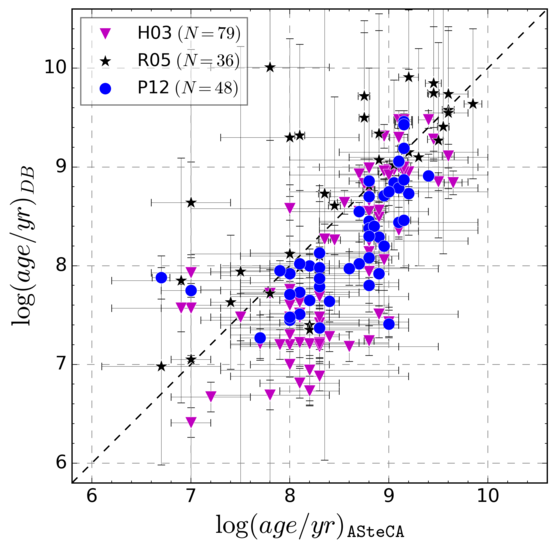}
\caption{Age comparison for integrated photometry DBs versus \texttt{ASteCA}.
$N$ is the number of clusters cross-matched in each DB.\@}
\label{fig:cross_match_ip_age}
\end{figure}

Our age and mass estimates are also compared with three DBs -- H03, R05, and P12
-- which used integrated photometry (see Table~\ref{tab:databases}). Only H03
and P12 obtained total mass values, these are analyzed in
Sect.~\ref{sssec:integ_photom_masses}.

H03 studied approximately 1000 clusters, 196 and 748 in the S/LMC, via $UBVR$
integrated photometry.
Ages were assigned based on the Starburst99
model~\citep{Leitherer_1999}, assuming metallicities, distance moduli, and
average $E_{B-V}$ values of (0.004, 0.008), (18.94, 18.48) mag, and 
(0.09, 0.13) mag, for the S/LMC, respectively.
Masses were derived from $M_V$ absolute magnitude and a mass-luminosity
relation.
This article represents, as far as we are aware, the largest published database
of MCs cluster masses to date.

R05 used two models -- GALEV~\citep{Anders_2003} and Starburst99 -- combined
with three metallicities -- (0.004, 0.008) and (0.001, 0.004, 0.008),
used in each model respectively -- to obtain five age estimates for 195 SMC
clusters. Reddening values were assigned using fixed age ranges, 
following~\cite{Harris_2004}.
We averaged all reddening-corrected ages for each matched cluster,
and assigned an error equal to the midpoint between the lowest and
highest error.

P12 used the same dataset from H03 to analyze 920 LMC clusters through their
MASSCLEAN$_{colors}$ and MASSCLEAN$_{age}$
packages~\citep{Popescu_2010a,Popescu_2010b}. Metallicities were fixed to
$z{=}0.008$, while reddening is taken from G10 when available, or fixed to
$E_{B-V}{=}0.13$ mag as done in H03. Ages and masses from duplicated entries in
the P12 sample are averaged in our analysis.


As seen in Fig.~\ref{fig:cross_match_ip_age}, H03 underestimates ages for
younger clusters. This effect was registered in~\citet[][see Fig. 1]
{de_Grijs_2006}, which the authors assigned to the photometry conversion done in
H03. Average $\log(age/yr)$ dispersion between H03 and \texttt{ASteCA}
is $0.44{\pm}0.56$ dex.
The same happens for P12 ages, albeit with a smaller $\log(age/yr)$ dispersion
of ${\sim}0.35{\pm}0.44$ dex. P12 compared their own age estimates with H03 (see
P12, Fig. 8). They find a systematic difference with H03, where MASSCLEAN
ages are larger for clusters with $\log(age/yr){<}8$ dex. In our case, most of
the clusters cross-matched with P12 are older than 8 dex, with P12 ages located
mostly below the identity line. This bias towards smaller age estimates by P12
is consistent with what was found in~\cite{Choudhury_2015}.
Contrary to the results found for H03 and P12, the R05 study slightly
underestimates ages compared to \texttt{ASteCA}, with a mean $\log(age/yr)$
dispersion of $-0.25{\pm}0.63$ dex. The standard deviation is the largest for
the three integrated photometry DBs. In R05 the authors mention the lack of
precision in their age measurements, due to the use of integrated colors, and
the lack of constrains for the metallicity.

Expectedly, the four isochrone fit studies analyzed previously show a more
balanced distribution of ages around the 1:1 relation, in contrast with the
integrated photometry DBs. Ages taken from integrated photometry studies are
known to be less accurate, and should be taken as a rather coarse approximation
to the true values.
As shown in P12, integrated colors present large scatters for all age values,
leading inevitably to degeneracies in the final solutions.
The added noise by contaminating field stars is also a key issue, as it is
very difficult to remove properly from integrated photometry data. A
single overly bright field star can substantially modify the observed
cluster's luminosity, leading to incorrect estimates of its
parameters~\citep{Baumgardt_2013,Piatti_2014_B88}.
A detailed analysis of some of the issues encountered by integrated photometry
studies, and the accuracy of their results, is presented in~\cite{Anders_2013}.


\subsubsection{Integrated photometry masses}
\label{sssec:integ_photom_masses}
%
%
%
%

\begin{figure*}
\centering
\includegraphics[width=2.\columnwidth]{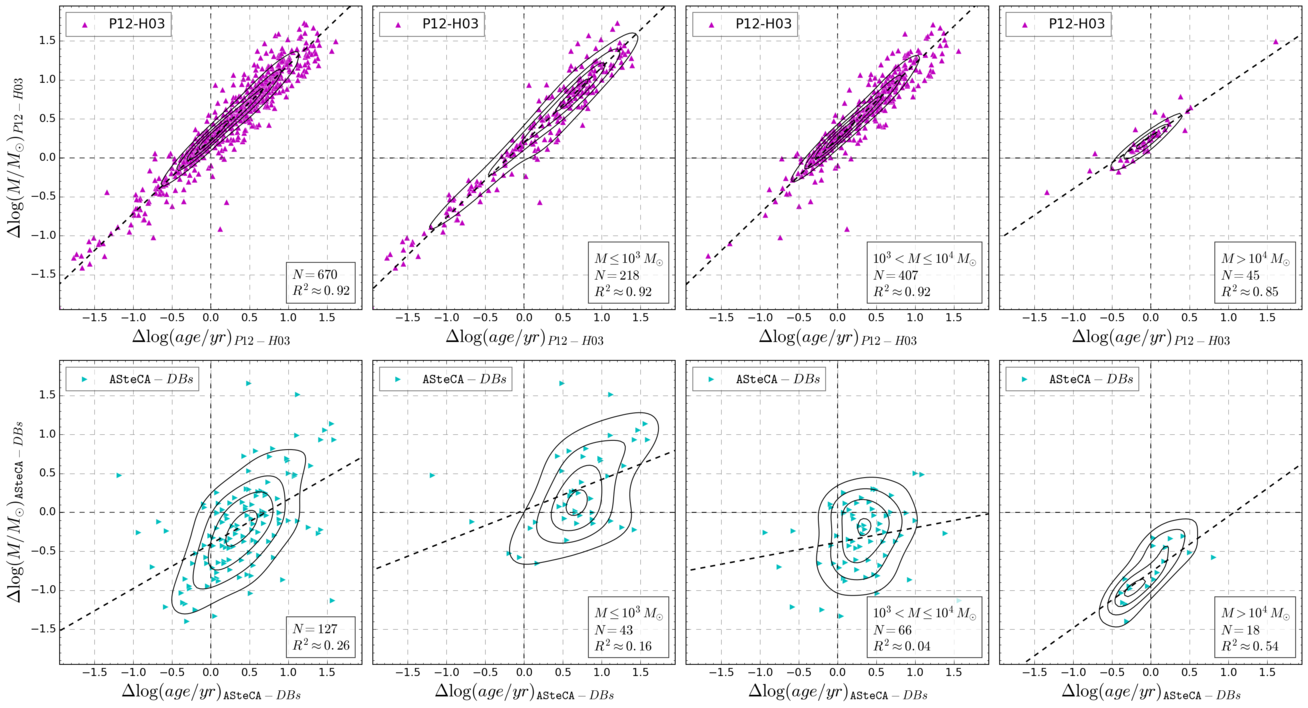}
\caption{\emph{Top row}: Differences plot $\log(age/yr)$ versus
$\log(M/M_{\odot})$, for 670 P12 and H03 cross-matched clusters, in the sense
P12 minus H03.
A 2-dimensional Gaussian kernel density estimate is shown as iso-density black
curves. The dashed line is the result of the best fit linear regression;
the $R^2$ coefficient is shown in the bottom left box.
The leftmost diagram shows all 670 clusters processed together, the remaining
diagrams are divided by mass ranges.
\emph{Bottom}: idem, for \texttt{ASteCA} versus DBs (where DBs represent the
mixed H03 and P12 sample) for the 127 cross-matched clusters.}
\label{fig:age_mass_corr}
\end{figure*}

Cross-matching the H03 and P12 DBs with a maximum search radius of 20
arcsec, results in 670 unique LMC clusters.
In the top row of Fig.~\ref{fig:age_mass_corr} we show the differences
in mass and age estimates for both DBs. The leftmost diagram contains all the
cross-matched clusters, while the other diagrams are sectioned by mass
ranges. The three mass regions show how the age-mass correlation -- an older
large cluster is incorrectly identified as a much younger and less massive one
or vice-versa -- strongly affects estimates between these DBs.\@
The mean value $\overline{\Delta M_{\log}}$ (logarithmic mass difference)
decreases from $0.8{\pm}0.5$, to $0.3{\pm}0.5$, to $-0.4{\pm}1$; in the
low, medium, and large mass regions defined in
Fig.~\ref{fig:age_mass_corr}, respectively. Masses go from being overestimated a
factor of ${\sim}6$ by P12 in the low mass region, to being underestimated by a
factor of ${\sim}2.5$ in the large mass region.
Where P12 estimates larger masses, it also assigns larger ages by more than 1.5
dex. Conversely, in the large average mass region, P12 ages can reach
values up to 3 dex lower than H03.
There are five clusters with H03 masses larger than
$2.5{\times}10^5\,M_{\odot}$ in this cross-matched sample, all incorrectly
assigned a low age and mass by P12. These five LMC globular clusters
are: NGC1916, NGC1835, NGC1786, NGC1754, and NGC1898. P12 assigns masses below
2000 $M_{\odot}$ in all cases.
%

Cross-matching the H03 and P12 DBs with our sample, results in a
set of 127 clusters. The age-mass positive correlation that was obvious for H03
and P12, is not present when we compare DBs masses and ages estimates with
those from \texttt{ASteCA}, as seen in the bottom row of
Fig.~\ref{fig:age_mass_corr}.
\begin{figure*}
\includegraphics[width=2.\columnwidth]{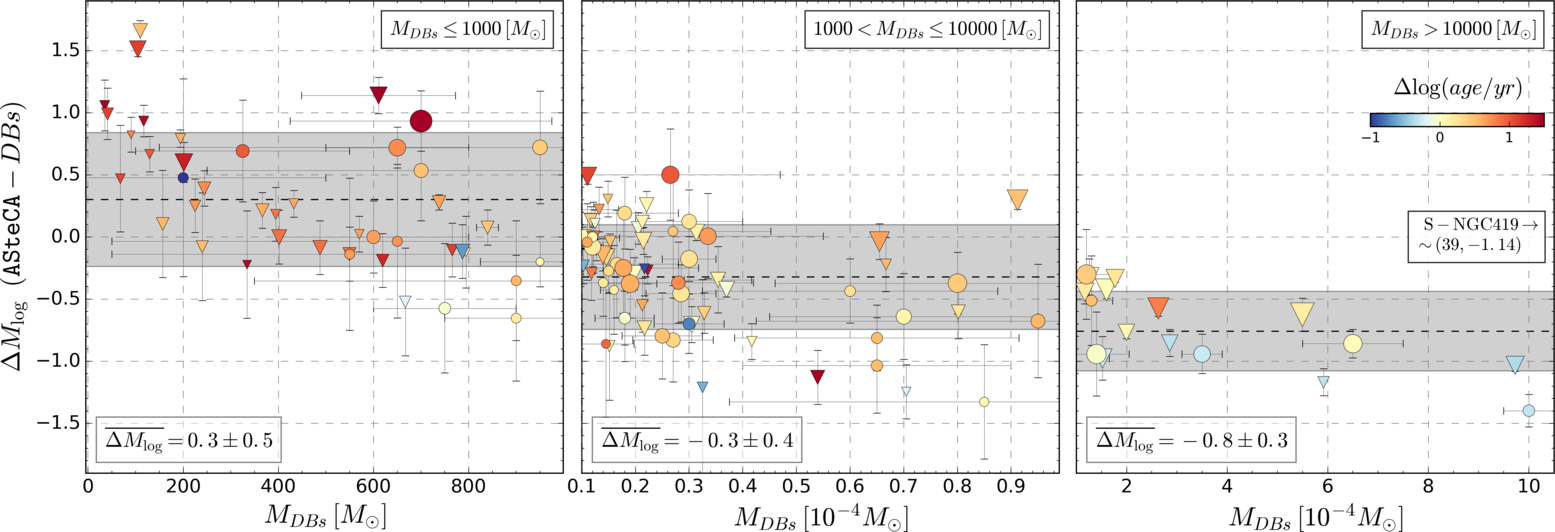}
\caption{\emph{Left}: BA mass plot, showing the differences between estimated
masses in the H03 and P12 DBs and the code, in the sense \texttt{ASteCA} minus
DB;\@ symbols as in Fig.~\ref{fig:cross_match_ip_age}.
Only DB masses ${\le}10^3\,M_{\odot}$ are shown here.
Colors are assigned according to the difference in $\log(age/yr)$ estimation of
each cluster (\texttt{ASteCA} minus DB, colorbar is shown in the right plot),
while sizes are proportional to the actual sizes in parsecs.
The horizontal dashed line ($\overline{\Delta M_{\log}}$) is the mean of the
combined logarithmic differences. The gray band is the ${\pm}1\sigma$ region for
this mean.
\emph{Center}: same as previous plot, now showing DB mass values in the range
$10^3-10^4\,M_{\odot}$.
\emph{Right}: same as previous plot, for DB mass values ${>}10^4\,M_{\odot}$.
}
\label{fig:cross_match_ip_mass}
\end{figure*}
In Fig.~\ref{fig:cross_match_ip_mass} we show masses for the 127 cross-matched
clusters in these DBs, versus their logarithmic mass differences.
%
The mean standard mass difference $\Delta M$ (${=}M_{\mathtt{ASteCA}}-M_{DBs}$)
is ${\sim}40{\pm}1700\,M_{\odot}$ for $M_{DBs} {<}5{\times}10^3\,M_{\odot}$,
and ${\sim-}400{\pm}2500\,M_{\odot}$ for $M_{DBs} {<}10^4\,M_{\odot}$; pointing
to a very reasonable scatter around the identity line.
Although the standard deviation values are somewhat large, this is expected for
a set of clusters in this mass range. As stated in~\cite{Baumgardt_2013} and
P12, clusters with $M{\lesssim}5{\times}10^3 - 10^4\,M_{\odot}$ tend to have
their estimated integrated photometry masses largely dominated by
stochastic processes.

We see in Fig.~\ref{fig:cross_match_ip_mass} that
the larger the DB mass, the larger the logarithmic difference with
\texttt{ASteCA}. Values for $\overline{\Delta M_{\log}}$ in the medium and
large-mass regions of Fig.~\ref{fig:cross_match_ip_mass} mean that DB masses are
between two and six times larger than \texttt{ASteCA} masses.
The most discrepant case is that of SMC-NGC419 -- left out of
the right plot for clarity -- which shows a mass of $3.9{\times}10^5\,M_{\odot}$
by H03, and $2.8{\times}10^4\,M_{\odot}$ by the code.

After exploring several possible processes that could induce this systematic
mass difference -- between our estimates and those from integrated photometry
studies--, we concluded that the responsible is the effect of stellar crowding
in our Washington photometry.
The likelihood defined in Eq.~\ref{eq:likelihood} is a binned statistic. This
means that stars not accounted for in the CMD will bias the mass estimation,
as it depends on the number of observed stars. The larger the number of stars
lost because of this process, the larger the underestimation of mass for a given
observed cluster.
To test the above hypothesis, we requested the HST data used
in~\cite{Goudfrooij_2014} to analyze NGC419. The ACS/WFC instrument -- from
where this data comes from -- has a much higher resolution compared to that from
our Washington photometry (0.05 arcsex/pixel versus 0.274 arcsec/pixel), so we
can expect a substantially lower percentage of stars lost.
The authors estimate a mass of ${\sim}2.4{\times}10^5\,M_{\odot}$ using
a~\cite{Salpeter_1955} IMF (which could be as low as
${\sim}1.5{\times}10^5\,M_{\odot}$ if a more recent IMF was used).
We analyzed this dataset with \texttt{ASteCA} fixing all parameters to the
values given in Goudfrooij et al., with the exception of the
mass which was left to vary up to $3{\times}10^5\,M_{\odot}$. A cut on $F555W
{=}23$ mag was imposed to minimize the impact of faint stars lost due to
crowding.
The mass found this way for NGC419 is $2.5{\times}10^5\,M_{\odot}$, a
very similar value to the average of H03 and Goudfrooij et al
($2.6{\times}10^5\,M_{\odot}$).
It is thus clear that stellar crowding in our low resolution photometry is the
responsible for the systematic difference in estimated masses.

Even though \texttt{ASteCA} corrects synthetic clusters using a
completeness function -- approximated from the observed cluster's LF--, this
only affects the faintest synthetic stars. For heavily crowded clusters 
(especially if they are observed with low resolution) this underestimates the
loss of brighter stars which intensifies as one moves closer to the
cluster's center~\citep{Mateo_1988}, hence underestimating its mass.
Future releases of the code will allow the user to input a manual
completeness function, ideally obtained through proper artificial star
tests~\citep[see e.g.,][]{Aparicio_Gallart_1995}.
Mass values obtained in the present study for massive clusters, should therefore
be considered lower estimates of their true value.


\section{Fundamental parameters in the analyzed database}
\label{sec:param-dist}

We present a summary of the distribution of the five fundamental parameters
obtained with \texttt{ASteCA}, for the 239 MCs clusters in our catalog.
A method is devised in Sect.~\ref{ssec:kde_method} to allow an unbiased analysis
of the estimated values, and the results presented in
Sect.~\ref{ssec:dist_ranges}. Reddenings, distances, and total masses do not
span sufficiently large ranges to account for the values found in the MCs. Their
distribution can thus only be though to characterize the state of those clusters
in this catalog.
On the other hand, metallicities and ages obtained cover a wide range in both
galaxies. Their distribution can be regarded as a representative randomized
sample of the cluster system in the MCs. Their relationship is studied in
Sect.\ref{ssec:amr}.

%



\subsection{Method}
\label{ssec:kde_method}

Histograms are used to derive a large number of properties in astrophysical
analysis, for example a galaxy's star formation history (SFH).
%
Their widespread use notwithstanding, the generation of a histogram is affected
by well known issues~\citep[see][]{Silverman_1986,Simonoff_1997}.
Different bin widths and anchor positions can make histograms built from the
same data look utterly dissimilar. In the worst cases, completely spurious
sub-structures may appear, leading the analysis towards erroneous conclusions.
We bypass these issues by constructing an adaptive (variable) Gaussian kernel
density estimate (KDE) in one and two dimensions, using the parameters' standard
deviations as bandwidth estimates. The formulas for both KDEs are:

\begin{equation}
KDE_{1D}(x) = \frac{1}{N\sqrt{2\pi}} \sum_{i=1}^N \frac{1}{\sigma_i}
e^{-\frac{(x-x_i)^2}{2\sigma_i^2}},
\label{eq:kde-1d}
\end{equation}

\begin{equation}
KDE_{2D}(x,y) = \frac{1}{2\pi N} \sum_{i=1}^N \frac{1}{\sigma_{xi}\sigma_{yi}}
e^{-\frac{1}{2} \left( \frac{(x-x_i)^2}{\sigma_{xi}^2} + 
\frac{(y-y_i)^2}{\sigma_{yi}^2} \right)},
\label{eq:kde-2d}
\end{equation}

\noindent where $N$ is the number of observed values, $x_i$ is the $ith$
observed value of parameter $x$, and $\sigma_{xi}$ its assigned standard
deviation (same for $y_i$ and $\sigma_{yi}$). The 1D version of these KDEs is
similar to the ``smoothed histogram'' used in the~\cite{Rafelski_2005} study of
SMC clusters.
Using standard deviations as bandwidth estimates means that the contribution to
the density map of parameters with large errors, will be smoothed (``spread
out'') over a large portion of the domain. Precise parameter values on the other
hand, will contribute to a much more narrow region.

Replacing one and two-dimensional histogram analysis with these KDEs has two
immediate benefits: a) it frees us from having to select an arbitrary bandwidth
value (the most important component of a KDE), and b) it naturally incorporates
errors obtained for each parameter into its probability density function.
%


\subsection{Distribution of parameters within the observed ranges}
\label{ssec:dist_ranges}

\begin{figure*}
\includegraphics[width=2.\columnwidth]{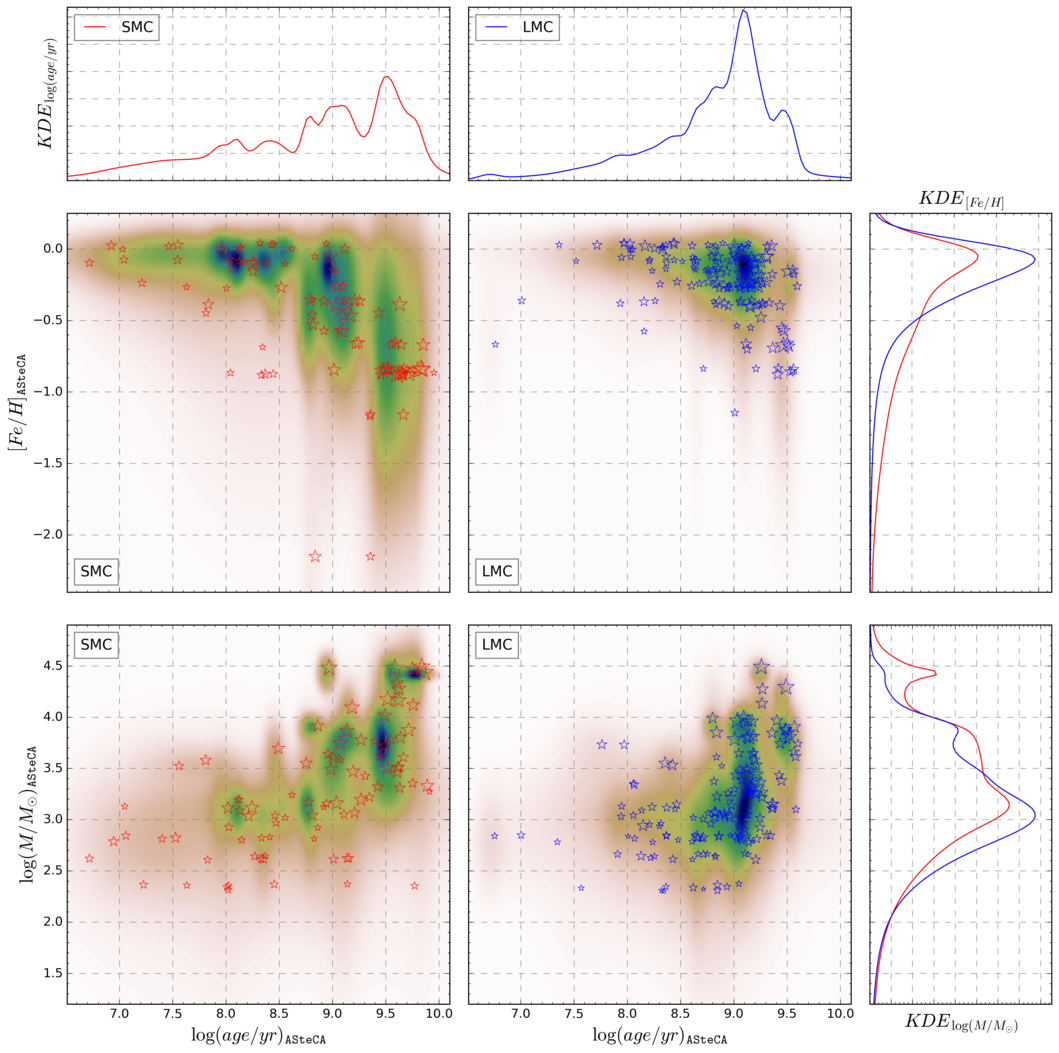}
\caption{Adaptive Gaussian KDEs for age, metallicity, and mass. Top and right
plots are 1D KDEs, center plots are 2D KDEs. Observed clusters are plotted as
red and blue stars for the S/LMC, respectively. Sizes are scaled according to
each cluster's radius. A small scatter is introduced for clarity.}
\label{fig:kde_fig0}
\end{figure*}

\begin{figure*}
\includegraphics[width=2.\columnwidth]{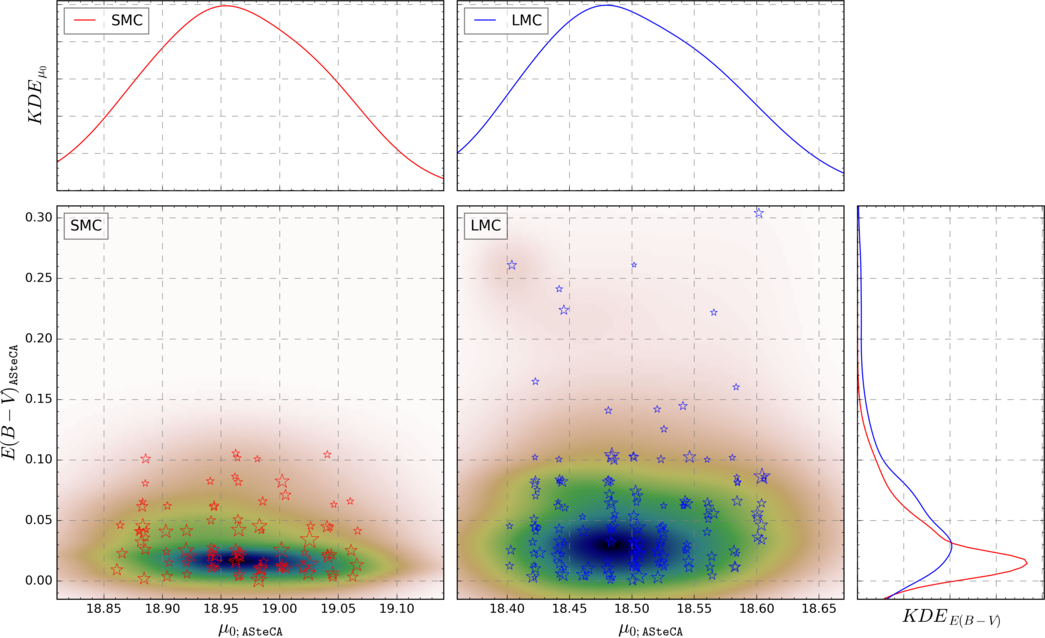}
\caption{Same as Fig.~\ref{fig:kde_fig0} for $E_{(B-V)}$ and distance
modulus.}
\label{fig:kde_fig1}
\end{figure*}

Figs.~\ref{fig:kde_fig0} and~\ref{fig:kde_fig1} show 1D and 2D density maps
constructed via Eqs.~\ref{eq:kde-1d} and~\ref{eq:kde-2d}.
%
%
A distinct period of cluster formation is visible in the LMC around ${\sim}$5
Gyr, which culminated ${\sim}$1.3 Gyrs ago. A similar but less
pronounced peak is seen for the SMC, with a drop in cluster formation around
${\sim}$2 Gyr.
Height differences between the SMC and LMC KDEs, is related to the relative
decline in cluster formation. While the LMC sharply drops to almost zero from
${\sim}$1 Gyr to present times, the SMC shows a softer descent with smaller
peaks around ${\sim}$250 Myr and ${\sim}$130 Myr.
The well known ``age gap'' in the LMC between 3--10 Gyrs~\citep{Balbinot_2010}
is present, visible as a marked drop in the $KDE_{\log(age/yr)}$ curve at
$\sim9.5$ dex.

The 2D KDE age-metallicity map shows how spread out these values are for
clusters in the SMC, compared to those in the LMC.\@ Although the SMC abundance
reaches substantially lower values, the right 1D KDE reveals [Fe/H] peaks
between 0 dex and -0.2 dex, for clusters in both Clouds.

The age-mass 2D map shows a clustering around younger ages and smaller masses
for the LMC.\@ The SMC cluster seen in the bottom right corner is HW42
($\alpha{=}1^h01^m08^s$, $\delta{=}-74^\circ04'25''$ [J2000.0]),
a small cluster (radius ${<}20$ pc) located close the the SMC's center. Though
its position in the map is somewhat anomalous, the 1$\sigma$ error in its age
and mass estimates could move it to
$\log(age/yr){\simeq}9.4$ and $\log(M/M_{\odot}){\simeq}2.6$. This cluster is
classified as a possible emissionless association by~\mbox{\cite{Bica_1995}}.
There is a tendency in both Clouds for the clusters' mass and size to
grow with estimated age, as expected~\citep[due to the mass-to-light ratio
increase with age; see][Sect. 4]{Popescu_2012}.

As seen in Fig.~\ref{fig:kde_fig1} (top), the 1D KDEs of the distance
moduli are well behaved and normal in their distribution.
A Gaussian fit to these curves results in best fit values of 18.96$\pm$0.08 mag
and 18.49$\pm$0.08 mag for the S/LMC.\@ Literature mean distances are thus
properly recovered.
Reddening values are much more concentrated in the sample of SMC clusters
around $E_{B-V}{\approx}0.015$ mag. LMC reddenings are dispersed below $E_{B-V}
{\approx}0.1$ mag, with a shallower peak located at ${\sim}0.03$ mag.
%


\subsection{Age-metallicity relation}
\label{ssec:amr}

A stellar system's age-metallicity relation (AMR) is an essential tool to learn
about its chemical enrichment evolution.
%
In~\cite{Piatti_2010_AMR} an AMR method was devised using age bins of different
sizes, to take age errors into account. It was applied to derive AMRs
in~\cite{Piatti_Geisler_2013}, and adapted to obtain star cluster frequency
distributions in~\cite{Piatti_2013_CF}.
We propose a new method based on the KDE technique described in
Sect.~\ref{sec:param-dist}, with a number of advantages over previous ones.
Mathematical details are given in Appendix~\ref{apdx:amr_description}, where
the method is applied to generate AMRs using literature age and metallicity
values.

\begin{figure}
\centering
\includegraphics[width=\hsize]{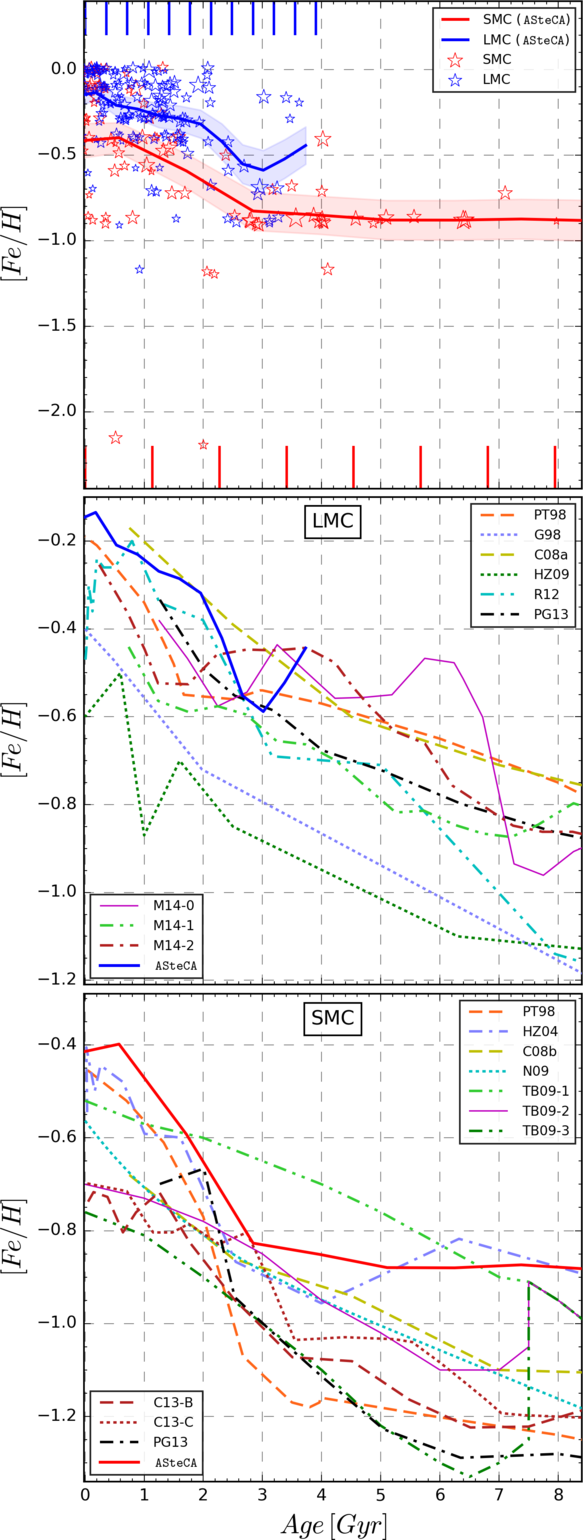}
\caption{\texttt{ASteCA}'s age-metallicity relation for the S/LMC (red/blue
solid lines). See text in Sect.~\ref{ssec:amr} for more details.}
\label{fig:amr}
\end{figure}

\texttt{ASteCA} AMRs for the S/LMC can be seen in Fig.~\ref{fig:amr} as red and
blue continuous lines, respectively. Stars show the position of all clusters in
our sample, with sizes scaled according to their radii.
Shaded regions represent the 1$\sigma$ standard deviations of the AMRs, spanning
a [Fe/H] width of ${\sim}0.2$ dex for the entire age range, for both Clouds.
The blue (top) and red (bottom) vertical segments in the top plot are the bin
edges determined for each age interval by Knuth's algorithm.
%
The final AMR functions are mostly unaffected by the chosen binning method.
Using Knuth's algorithm results in ${sim}11$ age intervals of widths between
0.35 and 1 Gyr, as seen in Fig.~\ref{fig:amr}. If instead we use 100 intervals
of ${\sim}0.1$ Gyr width, only the SMC curve is perturbed in the region
$\mathrm{Age}{<800}$ Myr where [Fe/H] values are raised by ${\sim}0.1$ dex.
%
%
If the two SMC clusters with very low metal abundances ([Fe/H]${<-}2$ dex) are
excluded from our data, the AMR moves upwards in the [Fe/H] axis by less than
0.05 dex, for ages below 500 Myr.

Several chemical evolution models and empirical AMRs are found in published
articles.
We show in Fig.~\ref{fig:amr} -- center and bottom plots -- AMR functions
presented in twelve other works.
These external studies constitute a representative sample of the different
methods and data used over the past twenty years:
\citet[][PT98; bursting models]{Pagel_1998}, \citet[][G98; closed-box model
with Holtzman SFH]{Geha_1998}, \citet[][HZ04]{Harris_2004}, \citet[][C08a;
average of four disk frames]{Carrera_2008_lmc}, \citet[][C08b; average of
thirteen frames]{Carrera_2008_smc}, \citet[][HZ09]{Harris_2009}, \citet[][N09;
5th degree polynomial fit to the AMRs of their three observed regions]
{Noel_2009}, \citet[][TB09; 1: no merger model, 2: equal mass merger, 3: one
to four merger]{Tsujimoto_2009}, \citet[][R12; four tiles average]{Rubele_2012},
\citet[][C13; B: Bologna, C: Cole]{Cignoni_2013}, \citet[][PG13]
{Piatti_Geisler_2013}, and \citet[][M14; 0: field LMC0, 1: field LMC1, 2: field
LMC2]{Meschin_2014}.
Details on how these AMRs were constructed can be consulted in each reference.
All of the above mentioned articles used field stars for the obtention
of their AMRs. This is, as far as we are aware, the first work were the AMR
function for both galaxies is derived entirely from observed star clusters.


The AMRs' trend coincides with what has already been found, namely that the
metallicity increases for younger ages (particularly below 3 Gyr). On average,
our AMR estimates are displaced slightly towards more metal rich values.
Most [Fe/H] values in the external studies are obtained using a solar
metallicity of $z_{\odot}{=}0.019$, while we used the more recent value
$z_{\odot}{=}0.0152$. As was shown in Sect.~\ref{ssec:lit-values}, this
difference means [Fe/H] estimates will be ${\sim}0.1$ dex smaller for external
studies.

For the LMC galaxy, Fig.~\ref{fig:amr} center plot, we see a marked
drop in metallicity from ${\sim-}0.45$ dex beginning around 3.8 Gyr, and
ending 3 Gyrs ago at ${\sim-}0.6$ dex. The M14-0 curve seems to reproduce this
behavior, but shifted ${\sim}0.8$ Gyr towards younger ages.
This high metallicity value of the LMC's AMR at its old end, is
caused by the three clusters with [Fe/H]${\approx-0.2}$ dex located beyond
${\sim}3.5$ Gyr; the oldest ages estimated by \texttt{ASteCA}. Without any older
clusters available in the LMC, it is hard to assess whether this is a
statistically significant feature of the AMR.
%
%
After the drop, there is a steep climb from 3 Gyr to 2 Gyr reaching almost 
[Fe/H]${\sim-}0.3$ dex, and then a sustained shallower increase up
to the estimated present day's metal content of ${\sim-}0.15$ dex.
Our average metallicity value for present day clusters coincides reasonably
well with the PT98 bursting model, which shows nonetheless a very different
rate of increase from 2 Gyr to present times. The C08a AMR, while lacking finer
details, provides a better match for this age range.
The HZ09 and G98 AMRs differ the most not only from the \texttt{ASteCA}
AMR, but from the rest of the group.

The SMC \texttt{ASteCA} AMR is shown along ten published AMRs in
Fig.~\ref{fig:amr}, bottom.
%
%
The peak around ${\sim}7.5$ Gyr predicted by TB09 in its two merger models 
(1:1, and 1:4 merger) is not visible in our AMR.\@
\texttt{ASteCA}'s AMR remains largely stable around a value of
[Fe/H]${\simeq-}0.9$ dex until approximately 3 Gyrs ago, where the rate of
growth increases considerably. From that point up to the present day, the
average metallicity for clusters in the SMC grows by ${\sim}0.4$ dex.
The abundance increase for ages ${<}3$ Gyrs, is only reproduced by the PT8
model, and the HZ04 function.
The PT98 model starts diverging from our AMR at ${\sim}2$ Gyr, until a gap of
${\sim}0.4$ dex in [Fe/H] is generated. In contrast, the HZ04 curve remains
much closer to our own throughout the entire age range. These two AMRs estimate
a present day metallicity very close to the [Fe/H]${\approx}-0.4$ dex value
estimated by \texttt{ASteCA}.


Overall, our AMRs can not be explained by any single model or empirical AMR
function, and are best reproduced by a combination of several. A similar result
was found in~\cite{Piatti_Geisler_2013} although their field stars AMRs are
significantly different from ours, mainly for the SMC.\@
It is important to remember that \texttt{ASteCA}'s AMRs are averaged over the
structure of both Magellanic Clouds. In Fig.~\ref{fig:ra-dec} we showed that our
set of clusters covers a large portion of the surface of these galaxies.
If more clusters where available so that the AMRs could be estimated by S/LMC
sectors, it is possible that different results would arise.


\section{Summary and conclusions}
\label{sec:summ-concl}

We presented an homogeneous catalog of 239 star clusters in the Large and Small
Magellanic Clouds, observed with the Washington photometric system. The clusters
span a wide range in metallicity and age, and are spatially distributed
throughout both galaxies.
The fundamental parameters metallicity, age, reddening, distance modulus, and
total mass were determined using the \texttt{ASteCA} package.
This tool allows the automated processing of a cluster's positional and
photometric data, resulting in estimates of both its structural and
intrinsic/extrinsic properties.
As shown in Paper I, the advantages of using \texttt{ASteCA} include
reproducible and objective results, along with a proper handling of the
uncertainties involved in the synthetic cluster matching process.
This permits the generation of a truly homogeneous catalog of observed
clusters, with their parameters fully recovered.
Our resulting catalog is complete for all the analyzed parameters, including
metallicity and mass, two properties often assumed or not obtained at all.

Internal errors show no biases present in our determination of fundamental
parameters, as seen in Sect.~\ref{sec:errors-fit}.
%
The analysis of our results in Sect.~\ref{ssec:lit-values}, demonstrate that the
assigned values for the clusters are in good agreement with published literature
which used the same Washington photometry.
Metallicity was the most discrepant parameter, with \texttt{ASteCA}'s
[Fe/H] values on average ${\sim}0.22$ dex larger than those present in the
literature. Half of this difference is due to the solar abundance used in
this work. The remaining ${\sim}0.1$ dex is explained by the confirmation
bias effect in most cluster studies, that will assume canonical [Fe/H]
values rather than derive them through statistically valid means.
We also compared our results with articles that used different photometric
systems, in Sect.~\ref{ssec:db-values}. While the age differences in
this case are somewhat larger, they can be mostly explained by effects outside
the code.

We performed in Sect.~\ref{sssec:integ_photom_masses} a detailed comparative
study of masses obtained through integrated photometry studies, with our own
estimates from CMD analysis.
Although for smaller clusters the estimation is reasonable, mass values
are systematically underestimated for larger
clusters, due to the effect of stellar crowding in our own photometry.

A method for deriving the distribution of any fundamental parameter -- or a
combination of two of them -- is presented in Sect.~\ref{ssec:kde_method}. It
takes into account the information contained by the uncertainties, often
excluded from the analysis. By relying on Gaussian kernels, it is robust and
independent of ad-hoc binning choices.
An age-metallicity relation is derived in Sect.~\ref{ssec:amr}, for
cluster systems in both galaxies. The AMRs generated can not be fully matched by
any model or empirical determination found in the recent literature.

We demonstrated that the \texttt{ASteCA} package is able to produce proper
estimations for the fundamental parameters of observed star clusters,
within the limitations imposed by the photometric data.
A necessary statistically valid error analysis
can be performed, thanks to its built-in bootstrap error assignment method.
The tool is proven capable of operating almost entirely unassisted, on large
databases of clusters. This is an increasingly essential feature
of any astrophysical analysis tool, given the growing importance of big data and
the necessity to conduct research on large astronomical data sets.


\begin{acknowledgements}
The authors are very much indebted with the anonymous referee for the helpful
comments and suggestions that contributed to greatly improve the manuscript.
GIP would like to thank the help and assistance provided throughout the
redaction of several portions of this work by: D. Hunter, A. E. Dolphin,
M. Rafeslski, D. Zaritsky, T. Palma, F. F. S. Maia, B. Popescu,
H. Baumgardt, J. C. Forte, and P. Goudfrooij.
This research has made use of the
VizieR\footnote{\url{http://vizier.u-strasbg.fr/viz-bin/VizieR}} catalogue
access tool, operated at CDS, Strasbourg, France~\citep{Ochsenbein_2000}.
This research has made use of
``Aladin sky atlas''\footnote{\url{http://aladin.u-strasbg.fr/}} developed at
CDS, Strasbourg Observatory, France~\citep{Bonnarel2000,Boch2014}.
This research has made use of NASA's Astrophysics Data
System\footnote{\url{http://www.adsabs.harvard.edu/}}.
This research made use of
the Python language v2.7\footnote{\url{http://www.python.org/}}~\citep
{vanRossum_1995},
and the following packages:
NumPy\footnote{\url{http://www.numpy.org/}}~\citep{vanDerWalt_2011};
SciPy\footnote{\url{http://www.scipy.org/}}~\citep{Jones_2001};
Astropy\footnote{\url{http://www.astropy.org/}}, a community-developed core Python
package for Astronomy \citep{Astropy_2013};
scikit-learn\footnote{\url{http://scikit-learn.org/}}~\citep{pedregosa_2011};
matplotlib\footnote{\url{http://matplotlib.org/}}~\citep{hunter_2007}.
This research made use of the Tool for OPerations on Catalogues And
Tables~\citep[TOPCAT,][]{Taylor_2005}\footnote{\url
{http://www.starlink.ac.uk/topcat/}}.
\end{acknowledgements}

\bibliographystyle{aa} 
\bibliography{biblio} 


\begin{appendix}

\section{Total cluster mass validation}
\label{apdx:mass_valid}

The likelihood used in this work (Eq.~\ref{eq:likelihood}) allows
to set the total cluster mass as a free parameter to be optimized.
To validate \texttt{ASteCA}'s mass recovery we processed 768 synthetic
clusters generated with the MASSCLEAN tool, 384 for each Magellanic Cloud.
These clusters imitate the metallicity, and age range for clusters in
both Clouds, with a large maximum mass. The distance and reddening parameters
were fixed; see table~\ref{tab:massclean-range}.
The process of generating a MASSCLEAN cluster was described in Paper I, Sect. 3.
It is worth noting that these clusters are affected by stellar crowding,
only in their faintest magnitudes~\citep[using a theoretical
completeness function similar to that presented in][]{Small_2013}. As such,
their mass estimations will not suffer from the systematic underestimation seen
in Sect.~\ref{sssec:integ_photom_masses}.
Each MASSCLEAN cluster had its $V$ vs ($B-V)$ CMD (in the $UBRIJHK$ photometric
system) analyzed by \texttt{ASteCA}.

\begin{table*}
\centering
\caption{Parameter values used to generate the set of 768 MASSCLEAN clusters.}
\label{tab:massclean-range}
\begin{tabular}{lcc}
\hline\hline
 Parameter & Values & N\\
\hline
z & 0.001, 0.004, 0.015, 0.03 & 4\\
$\log\mathrm{(age/yr)}$ & 7, 7.2, 7.5, 7.7, 8, 8.2, 8.5, 8.7, 9, 9.2, 9.5, 9.7 &
12\\
$\mu_0$ & 18.9 (SMC), 18.5 (LMC) & 2\\
$E_{B-V}$ & 0.1 & 1\\
Mass ($M_{\odot}$) & 500, 1000, 5000, 10000, 25000, 50000, 100000, 250000 & 8\\
\hline
\end{tabular}
\end{table*}

\begin{figure*}
\includegraphics[width=2.\columnwidth]{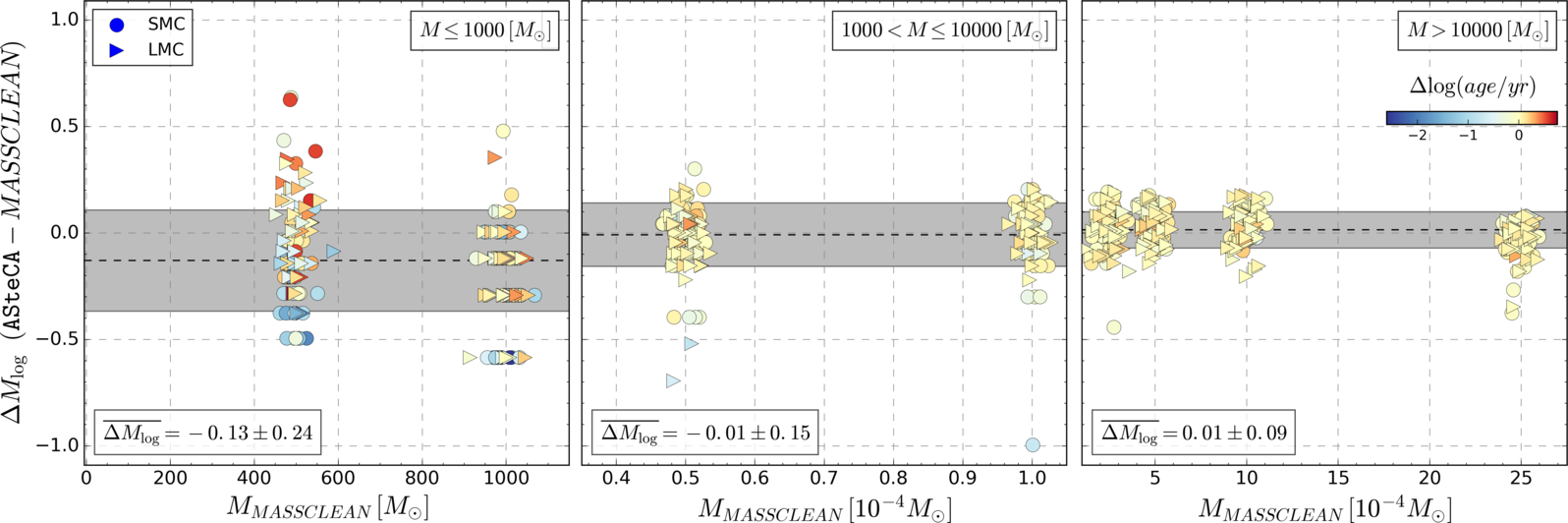}
\caption{Recovered masses by \texttt{ASteCA} for the 768 MASSCLEAN clusters.
Logarithmic mass differences $\Delta M_{\log}$ are obtained in the sense
\texttt{ASteCA} minus MASSCLEAN, and shown in the y-axis. MASSCLEAN masses in
the x-axis are perturbed with a small random scatter.}
\label{fig:massclean_mass}
\end{figure*}

In Fig~\ref{fig:massclean_mass} we show the masses recovered for the 768
MASSCLEAN clusters. The x-axis displays the true MASSCLEAN mass
values. The y-axis shows the logarithmic mass differences, in the sense \texttt
{ASteCA} minus MASSCLEAN.\@
Colors follow the differences in $\log(age/yr)$ (\texttt{ASteCA} minus
MASSCLEAN), shown in the right plot colorbar.
Average age differences for each mass region are: 
$-0.3{\pm}0.6$ dex ($\overline{M}{\le}1000\,[M_{\odot}]$),
$-0.05{\pm}0.19$ dex ($1000{<}\overline{M}{\le}10000\,[M_{\odot}]$),
$-0.01{\pm}0.13$ dex ($\overline{M}{>}10000\,[M_{\odot}]$).
As expected, clusters with larger masses have their ages more accurately
recovered. On average, the difference between \texttt{ASteCA} (estimated) minus
MASSCLEAN (true) logarithmic ages in the full mass range is ${\sim}-0.1{\pm}0.4$
dex.
%
Gray bands represent the mean and standard deviation for the logarithmic
mass differences, $\overline{\Delta M_{\log}}$.
%
For low mass clusters -- $500\,M_{\odot}$ or $1000\,M_{\odot}$ -- the
code assigns masses in a range between ${\sim}200{-}3000$ $M_{\odot}$.
In this region \texttt{ASteCA} underestimates clusters' masses by
${\sim}200\,M_{\odot}$. This effect is tied to an improper age estimation, where
\texttt{ASteCA} incorrectly assigns younger ages to scarcely populated clusters,
and compensates the low number of stars by decreasing the total mass. Such an
issue is not unexpected for very low mass clusters.

Table~\ref{tab:corr-matr} shows the correlation matrix between the five cluster
parameters. We see the usual correlations appear (age-metallicity,
metallicity-distance, age-reddening, etc.), as found in Paper I (Table 3).
Total mass shows a small positive correlation with the distance modulus.
When distance is overestimated, the matched synthetic CMD will contain fewer low
mass stars due to the magnitude limit. The likelihood will compensate this loss
by increasing its mass.

\begin{table*}
\centering
\caption{Correlation matrix for parameter deltas, defined for each cluster in
the sense \texttt{ASteCA} minus MASSCLEAN.}
\label{tab:corr-matr}
\begin{tabular}{lccccc}
\hline
\hline\\[-1.85ex]
$\Delta param$ & $\Delta z$ & $\Delta \log(age/yr)$ & $\Delta \mu$ &
$\Delta E_{B-V}$ & $\Delta M$ \\
\hline\\[-1.85ex]
$\Delta z$            &  1. & -0.36 & 0.24  & -0.15 & 0.03 \\
$\Delta \log(age/yr)$ &  -- & 1.    & -0.15 & -0.28 & 0.01 \\
$\Delta \mu$          &  -- & --    & 1.    & 0.05  & 0.13 \\
$\Delta E_{B-V}$      &  -- & --    & --    & 1.    & 0.0\\
$\Delta M$            &  -- & --    & --    & --    & 1. \\
\hline
\end{tabular}
\end{table*}



\subsection{Metallicity estimation for different mass values}
\label{apdx:ssec:metallicity}

\begin{figure*}
\includegraphics[width=2.\columnwidth]{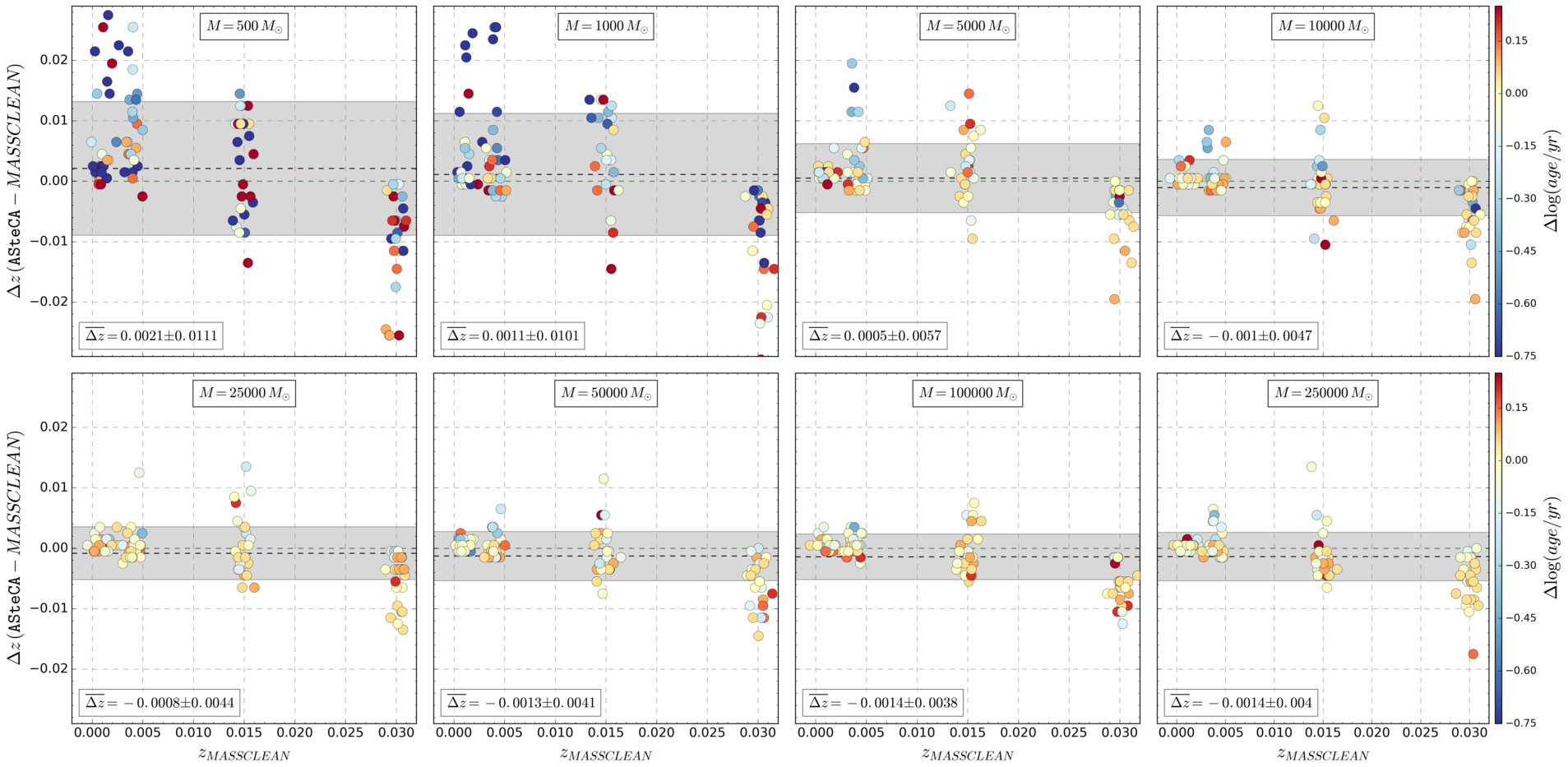}
\caption{\texttt{ASteCA} metallicity estimates for each mass used to generate
MASSCLEAN clusters. Colors are associated to $\log(age/yr)$ differences, shown
in the colorbars to the right. The green dashed horizontal line is the $\Delta
[Fe/H]{=}0$ line, shown as reference.}
\label{fig:massclean_z}
\end{figure*}

The metallicity ($z$) estimated for the MASSCLEAN set is displayed in
Fig.~\ref{fig:massclean_z}, where tendencies are visible.
First, as the cluster's mass grows so does the accuracy of the
metallicity estimates. Although the average difference between true and
estimated values remains close to $\overline{\Delta z}{\approx}0.001$ for the
entire mass range -- this is expected, as $z{=}0.001$ is the step used by
\texttt{ASteCA} --, its standard deviation drops from ${\sim}0.01$ to $0.004$
for the more massive clusters.
%
%
Most of the poorest solutions obtained by \texttt{ASteCA} -- those with
$|\Delta z|{>}0.01$ dex -- are associated to low mass scarcely populated
clusters, with ${\sim}40$ true member stars on average (from two up to a
hundred) in their analyzed CMDs. This poor solutions set is composed of 91
clusters -- ${\sim}12\%$ of the sample -- 58 of which have
$M{\le}1000\,M_{\odot}$.
Of these 58 low mass clusters, 38 are assigned younger ages by the code due to
an improper field star decontamination process (an expected issue when the
number of true members is very low).
Of the 82 clusters with the worst age estimates by \texttt{ASteCA} --
$|\Delta\log(age/yr)|{\ge}0.5$ dex, ${\sim}11\%$ of the MASSCLEAN sample --
${\sim}90\%$ (73) are clusters with $M{\le}1000\,M_{\odot}$.
%
Leaving out these 82 clusters, the average difference in $z$
for the entire mass range is ${\sim}0.0008{\pm}0.006$ dex; a rather
small difference with reasonable dispersion.

The second tendency is the overestimation of $z$ for the lowest metallicities,
and its underestimation for the largest ones.
A balanced distribution around $\Delta z{=}0$ line is mostly seen for
abundances in the middle portion of the analyzed range. This trend is more
noticeable for lower masses, but can be found for all mass values.
This is a statistical artifact that arises due to the necessarily limited
metallicity range analyzed. For clusters with the lowest metal contents ($z
{=}0.001$), \texttt{ASteCA} can only assign equal or larger metallicities since
negative $z$ values are not possible.
Equivalently, for clusters with the largest abundances ($z{=}0.03$) the code
can only associate equal or lower metallicities because of the upper
$z$ limit used by \texttt{ASteCA}, which is precisely $z{=}0.03$.
This ``bias'' could be avoided for large metallicity clusters, by
increasing the $z$ range upper limit. It can not be avoided for the lowest metal
abundances.\\

\begin{figure*}
\includegraphics[width=2.\columnwidth]{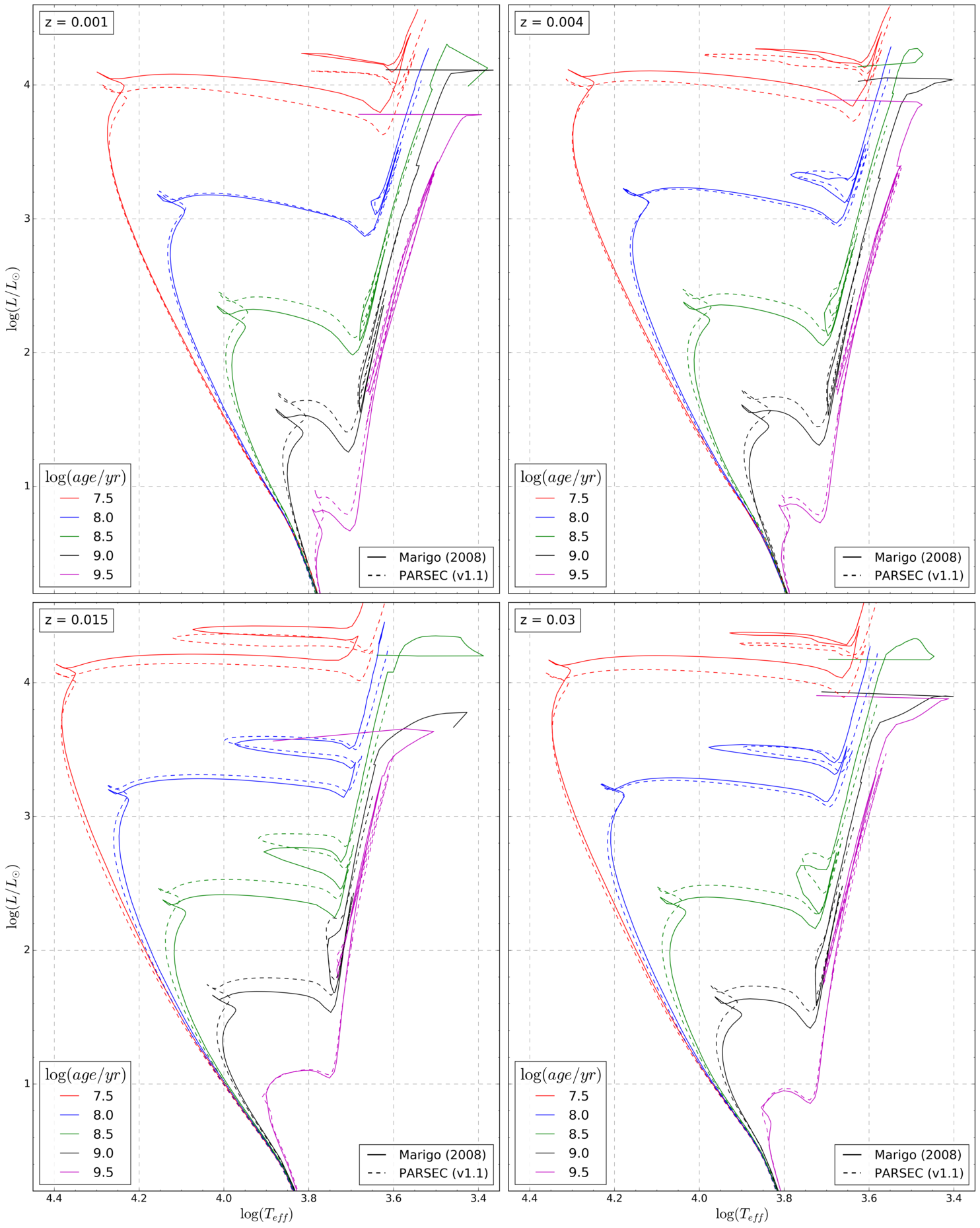}
\caption{\cite{Marigo_2008} versus PARSEC~\citep{Bressan_2012} isochrones, for
different metallicities and ages.}
\label{fig:marig_parsec}
\end{figure*}

An external source of errors also needs to be taken into account when analyzing
\texttt{ASteCA}'s metallicity (and age) estimates, for MASSCLEAN clusters.
This is the intrinsic differences between~\cite{Marigo_2008} isochrones --
used to generate MASSCLEAN clusters -- and PARSEC~\citep{Bressan_2012}
isochrones -- used by \texttt{ASteCA} to find the optimal fundamental
parameters. These differences are a source of error in the matching
process that is not straightforward to quantify.
The two sets of tracks have non-negligible dissimilarities beyond the turn-off
points, for most of the age range where they can be produced. This can be seen
in Fig.~\ref{fig:marig_parsec}, where isochrones from both sets are compared for
five different $\log(age/yr)$ values from 7.5 to 9.5 dex.
For ages up to 8 dex, PARSEC isochrones present a turn-off point located at
lower $\log(L/L_{\odot})$ values, particularly for lower metallicities. This
causes a shift in the more evolved portions of the isochrone, displacing the
Marigo isochrones towards larger $\log(L/L_{\odot})$ values.
Beyond that age this effects reverses, and PARSEC isochrones are now lifted
above the Marigo tracks. Given the many known correlations between fundamental
parameters it is not easy to predict how the matching algorithm will resolve
such instances.

\section{Outliers}
\label{apdx:outliers}

Ten of the analyzed clusters in this work -- ${\sim}4\%$ of the set -- show age
differences with the literature $\Delta\log(age/yr){>}0.5$.
Such a large age difference translates into two very dissimilar isochrones
fitted to the same coeval star sequence, which makes this sub-sample of
clusters stand out.
For these ``outliers'' no configuration of the DA plus the employed binning
methods could be found, that resulted in synthetic CMD matches with age values
close to those found in the literature.
%
All clusters in the outliers sample had smaller ages assigned by the code,
compared to the literature, see Table~~\ref{tab:outliers}. These
differences go from 0.55 dex up to 1.6 dex in the most extreme case of
LMC-KMHK975.

\begin{table}
\centering
\caption{Clusters with large age differences between the literature
and values found by the code (``outliers'').
Equatorial coordinates are expressed in degrees for the $J2000.0$ epoch.
Ages are given as $\log(age/yr)$ for literature (L) and \texttt{ASteCA} (A).
The difference between both estimates (L-A) is given in the last column as
$\Delta$.}
\label{tab:outliers}
\begin{tabular}{lccccc}
\hline\hline
Cluster & $\alpha(^\circ)$ & $\delta(^\circ)$ & L & A & $\Delta$\\
\hline
L-KMHK975 & 82.49583 & -67.87889 & 8.30 & 6.70 & 1.60\\
L-SL579 & 83.55417 & -67.85639 & 8.15 & 7.00 & 1.15\\
L-BSDL631 & 76.64167 & -68.42722 & 8.35 & 7.50 & 0.85\\
L-KMHK979 & 82.41250 & -70.98389 & 7.90 & 7.30 & 0.60\\
L-H88-316 & 85.41250 & -69.22944 & 8.25 & 7.70 & 0.55\\
\\[-1.85ex]
S-L35 & 12.00417 & -73.48611 & 8.34 & 6.90 & 1.44\\
S-H86-188 & 15.05833 & -72.45833 & 8.10 & 6.70 & 1.40\\
S-L39 & 12.32500 & -73.37167 & 8.05 & 7.00 & 1.05\\
S-B134 & 17.25417 & -73.20667 & 8.15 & 7.20 & 0.95\\
S-K47 & 15.79583 & -72.27361 & 7.90 & 7.00 & 0.90\\
\hline
\end{tabular}
\end{table}

\begin{figure*}
\includegraphics[width=2.\columnwidth]{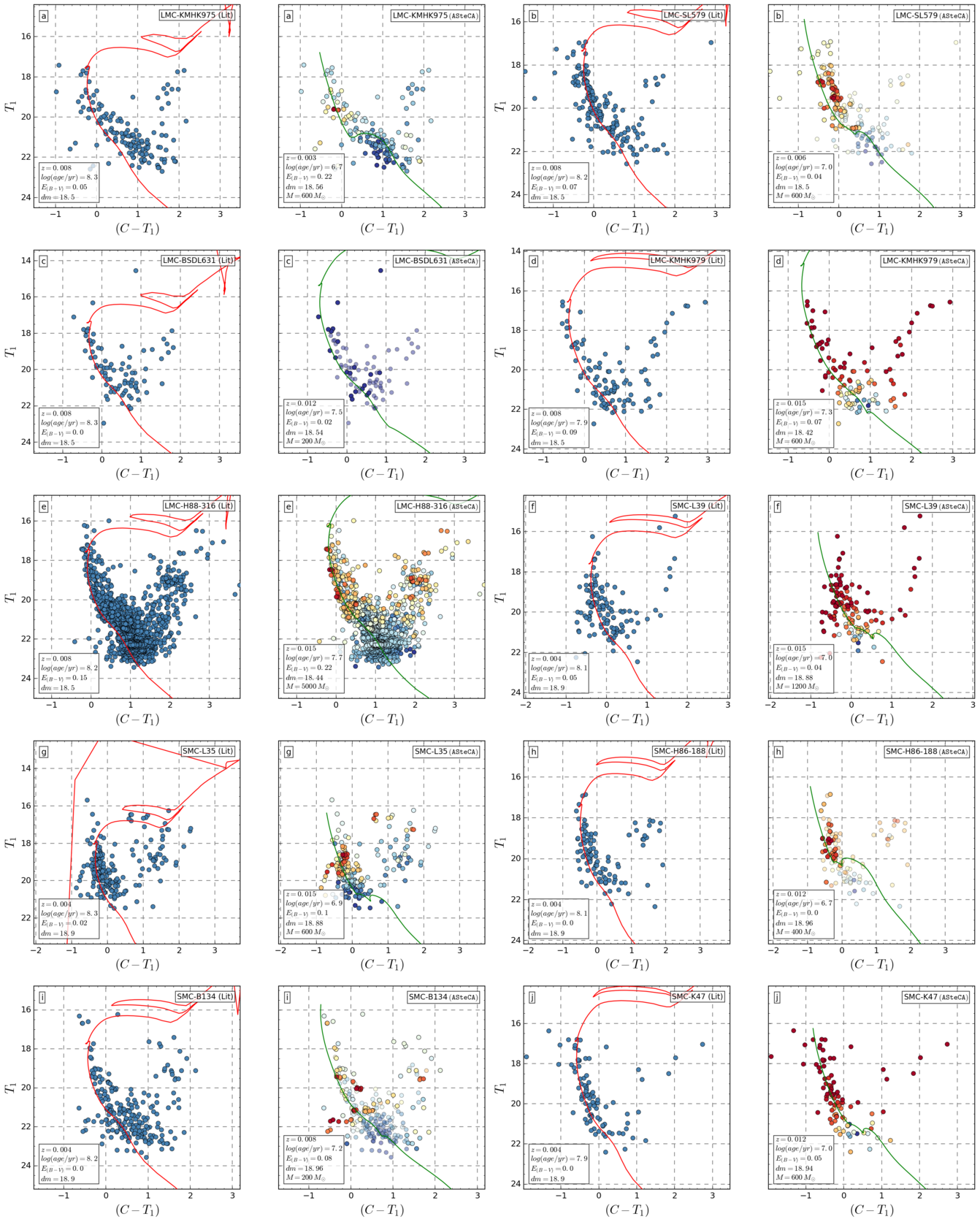}
\caption{CMDs for the outliers set. Description of the plots in the main
text of the section.}
\label{fig:outliers0}
\end{figure*}

In Fig.~\ref{fig:outliers0} CMDs for these clusters are plotted, two per row.
Each CMD pair shows the cluster region with the literature isochrone fit (left),
and the best match isochrone found by \texttt{ASteCA} (right).
%
Colors in the right CMD correspond to the MPs assigned by the DA, while
semi-transparent stars are those removed by the cell-by-cell density based
cleaning algorithm (see Sect.\ref{ssec:dencontamination}).
%

For most of the outliers, the same process is identified as the main
cause responsible for the observed age differences.
While the literature by-eye isochrone fit aligns the brighter
part of the cluster's sequence with the turn off point of a an older
isochrone, \texttt{ASteCA} decides instead that this is the top portion of a
much younger isochrone with no discernible turn off.
The statistical mismatch due to the removal of low mass stars by the DA
-- discussed in Sect.~\ref{ssec:lit-values} -- can also be seen to affect some
of the fits. In particular, SMC clusters SL579 and H86-188 show signs of
this effect in the best synthetic CMD match selected by \texttt{ASteCA}
(see Fig.~\ref{fig:outliers0}, CMDs \emph{b} and \emph{h}).

These age estimates could be brought closer to literature values, if a more
restrictive age range was used (e.g.: a minimum value of $\log(age/yr){=}7.5$
dex instead of 6 dex as used in this work).
Lacking external evidence to substantiate this a priori restriction, we choose
to keep the values obtained by \texttt{ASteCA}, with this section acting as a
cautionary note.

\section{Color-magnitude diagrams for the P99, P00, C06, and G10 databases}
\label{apdx:databases}

\begin{figure*}
\includegraphics[width=2.\columnwidth]{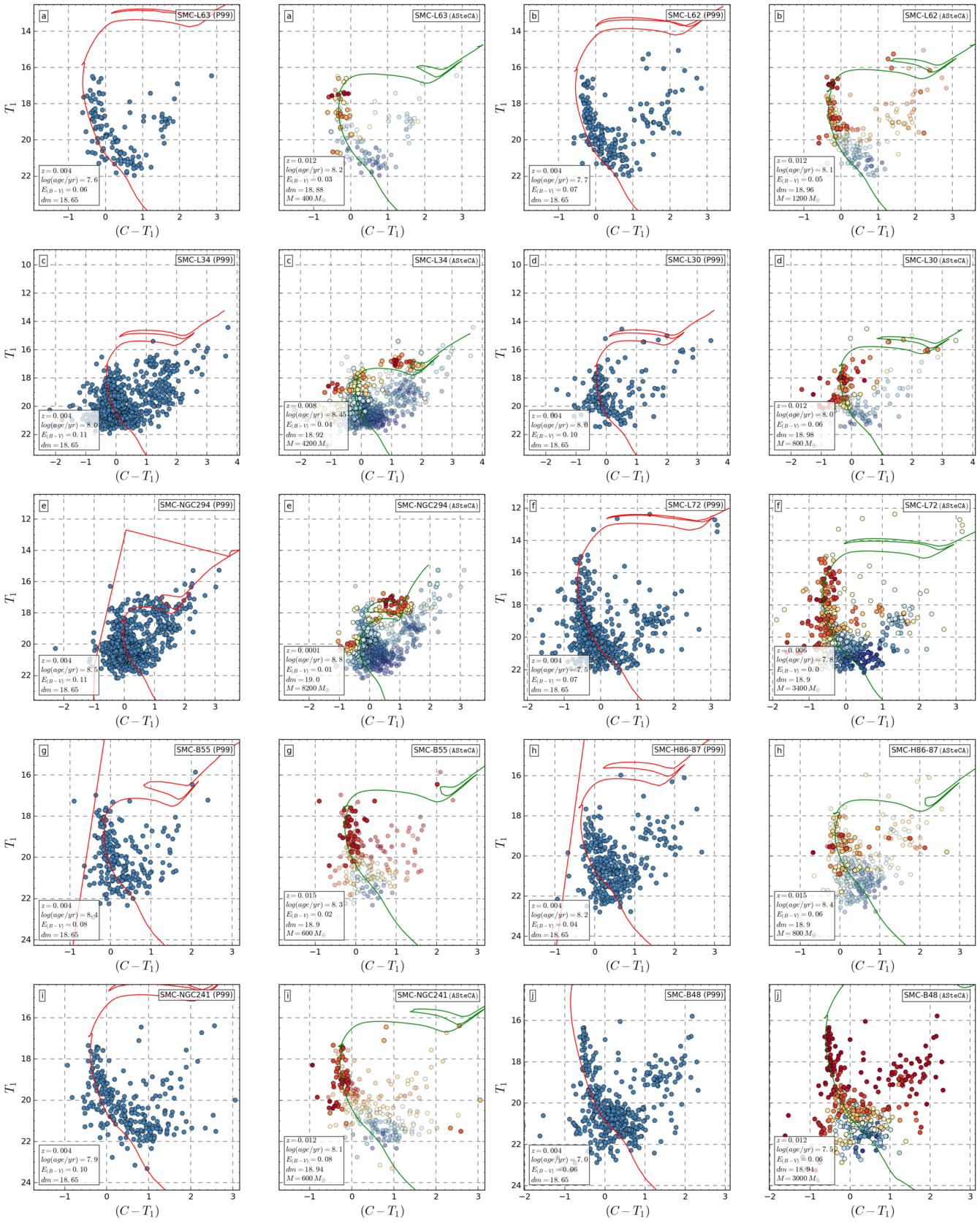}
\caption{CMDs for the P99 database.}
\label{fig:DBs_P99_0}
\end{figure*}

\begin{figure*}
\includegraphics[width=2.\columnwidth]{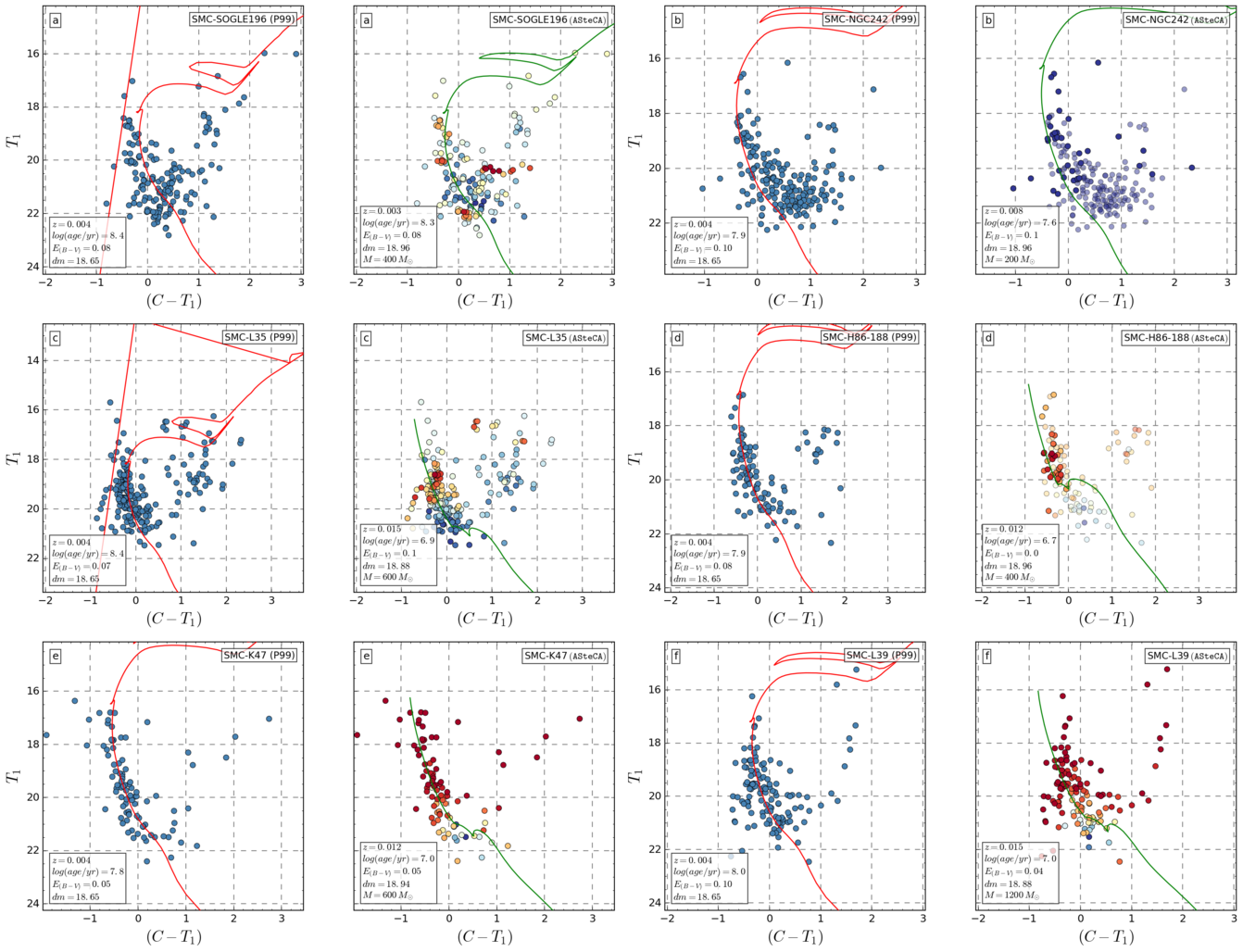}
\caption{CMDs for the P99 database.}
\label{fig:DBs_P99_1}
\end{figure*}

\begin{figure*}
\includegraphics[width=2.\columnwidth]{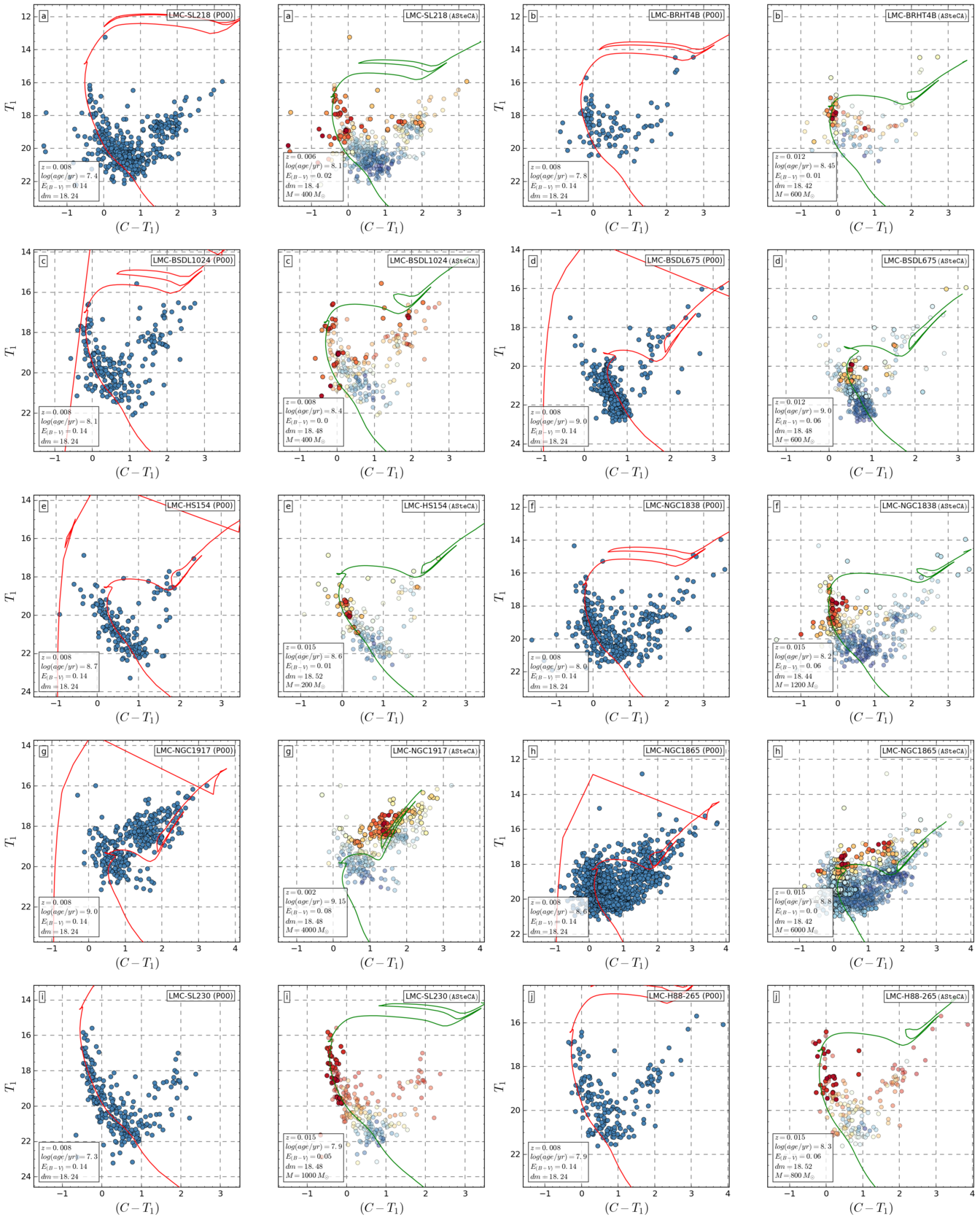}
\caption{CMDs for the P00 database.}
\label{fig:DBs_P00_0}
\end{figure*}

\begin{figure*}
\includegraphics[width=2.\columnwidth]{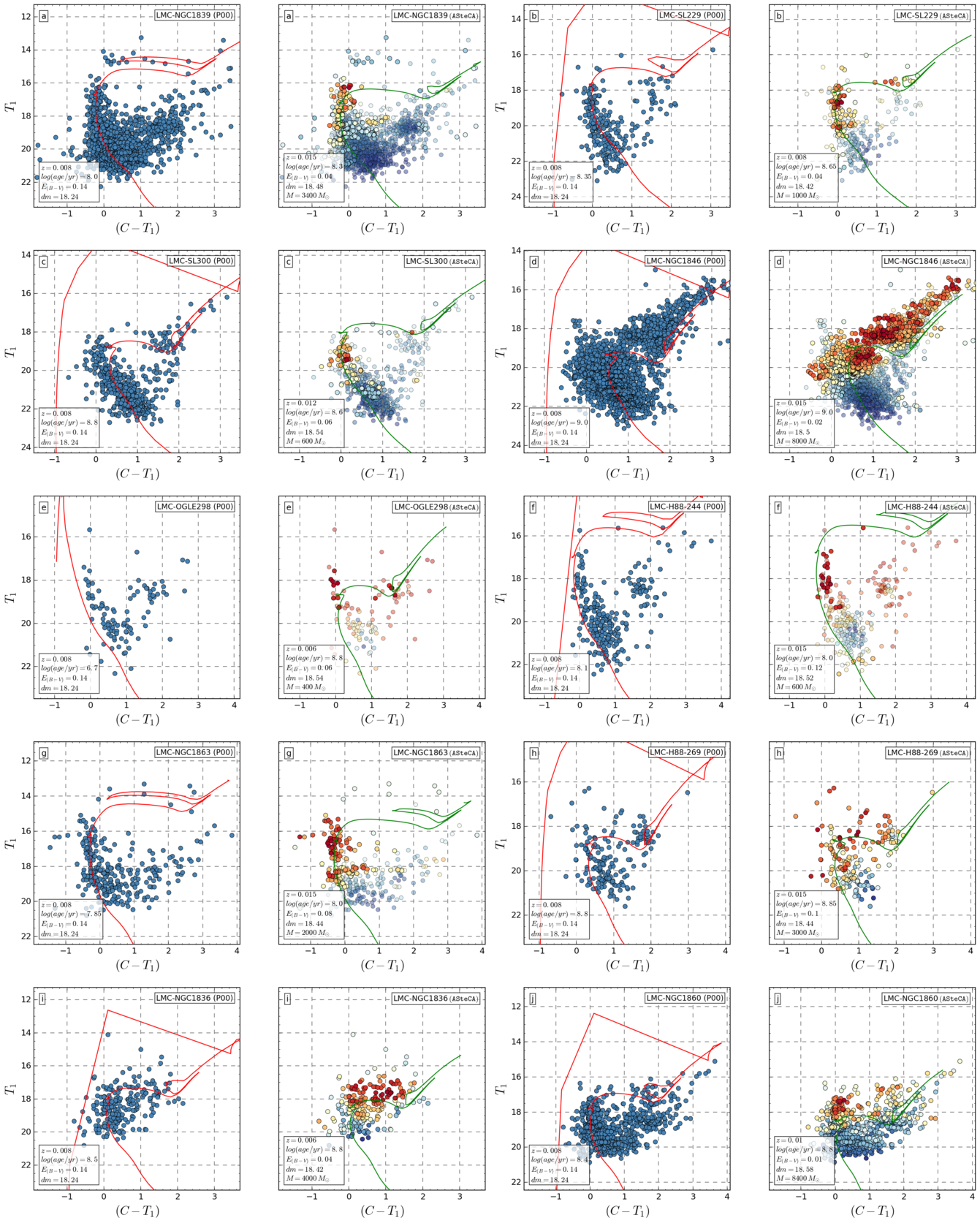}
\caption{CMDs for the P00 database.}
\label{fig:DBs_P00_1}
\end{figure*}

\begin{figure*}
\includegraphics[width=2.\columnwidth]{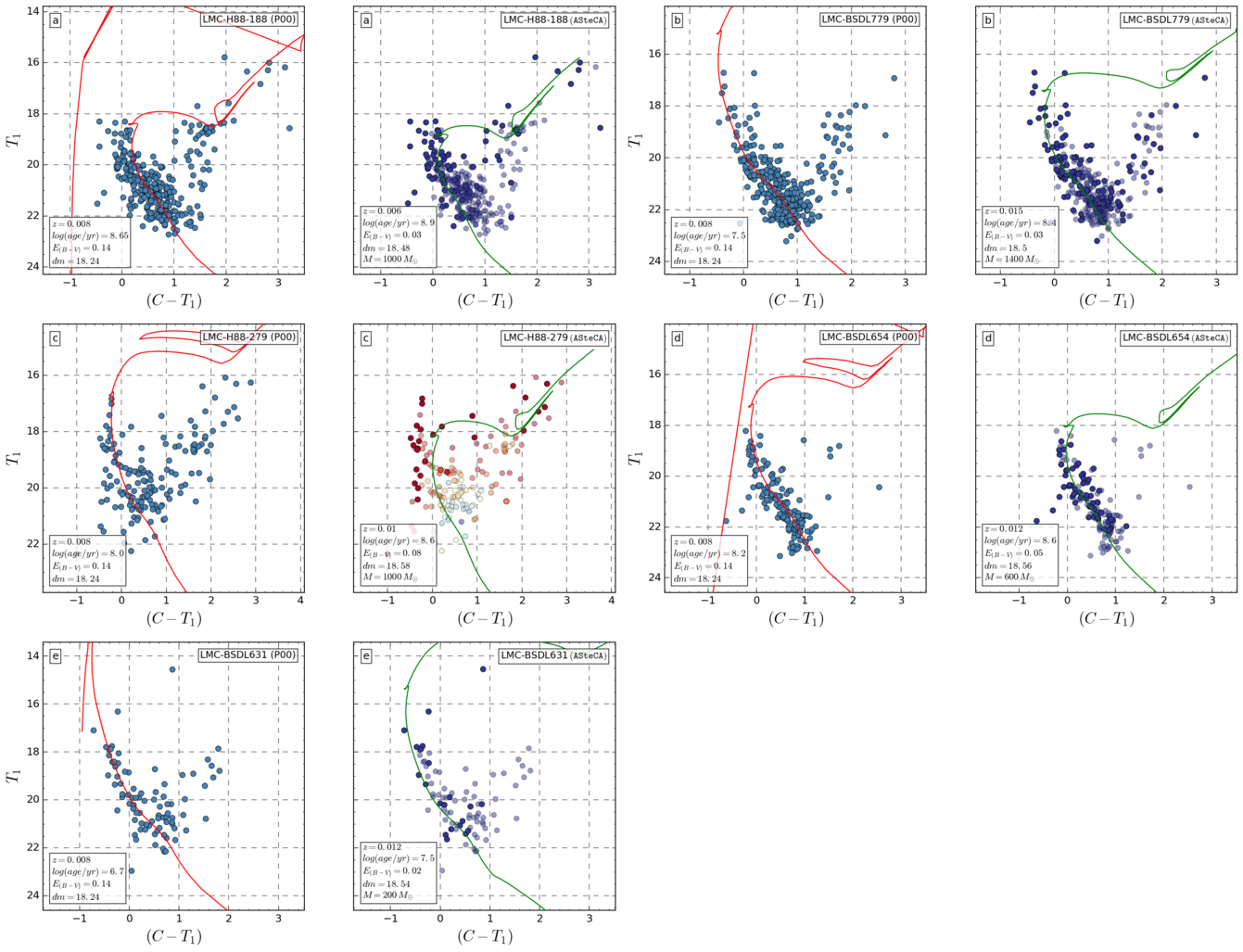}
\caption{CMDs for the P00 database.}
\label{fig:DBs_P00_2}
\end{figure*}

\begin{figure*}
\includegraphics[width=2.\columnwidth]{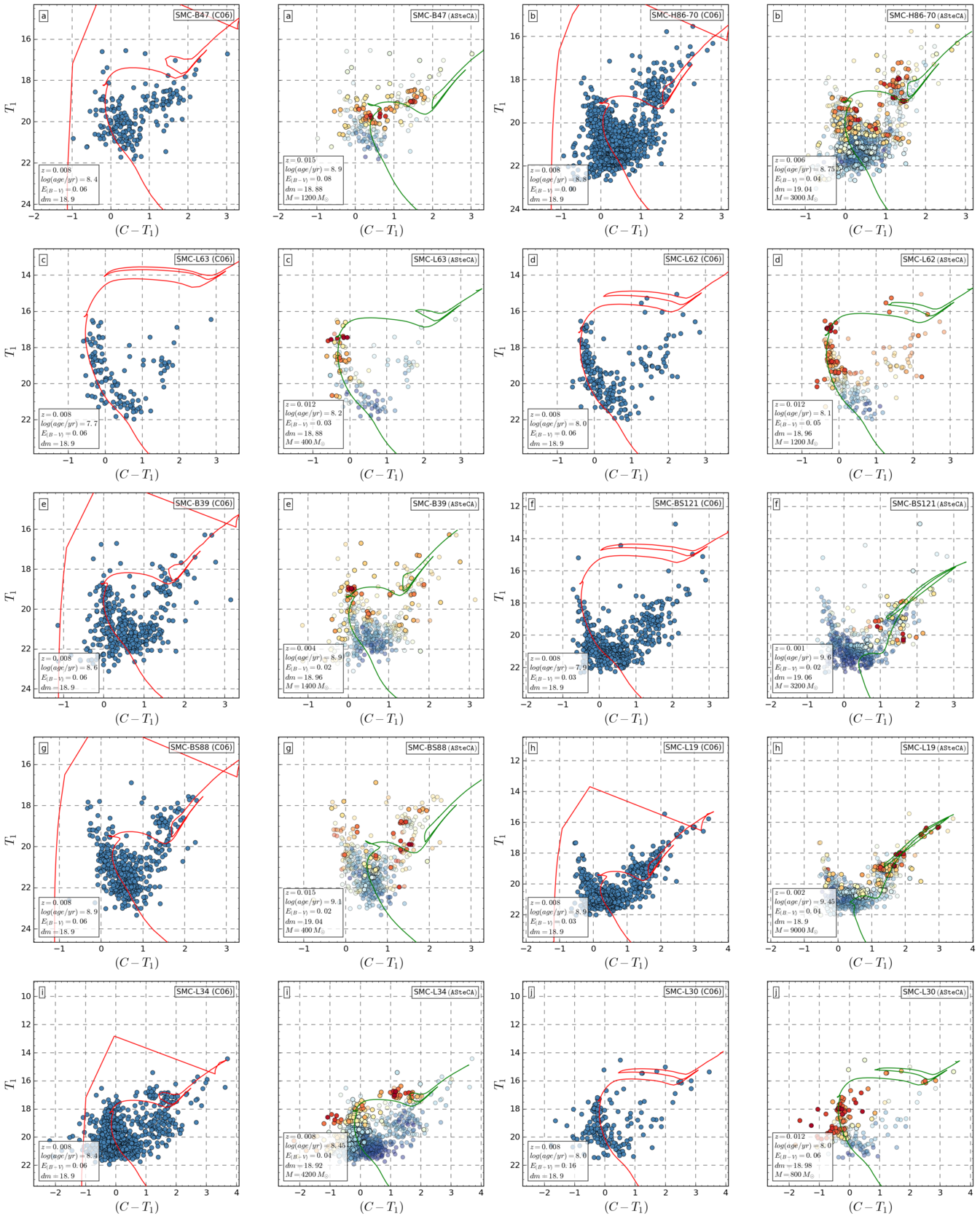}
\caption{CMDs for the C06 database.}
\label{fig:DBs_C06_0}
\end{figure*}

\begin{figure*}
\includegraphics[width=2.\columnwidth]{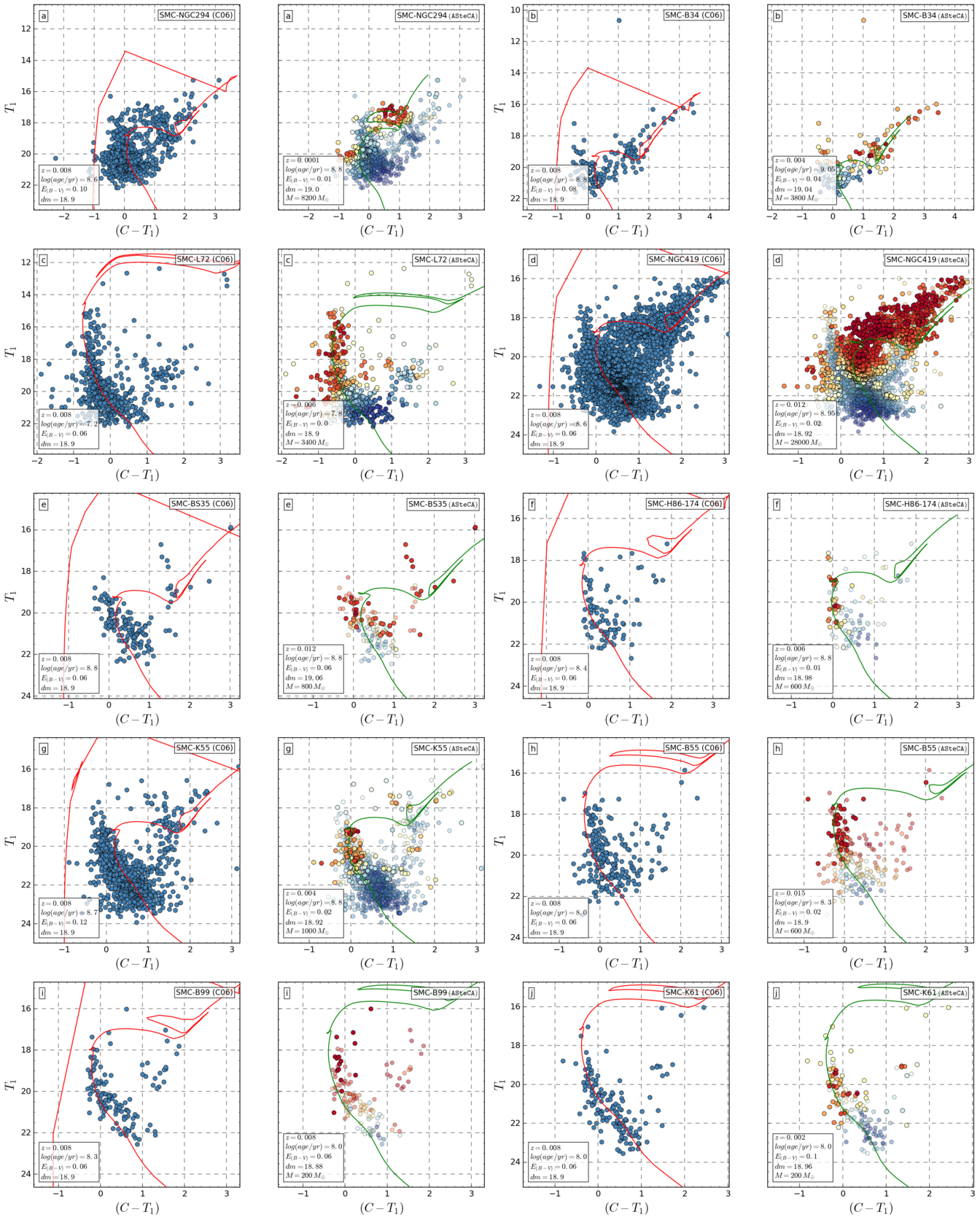}
\caption{CMDs for the C06 database.}
\label{fig:DBs_C06_1}
\end{figure*}

\begin{figure*}
\includegraphics[width=2.\columnwidth]{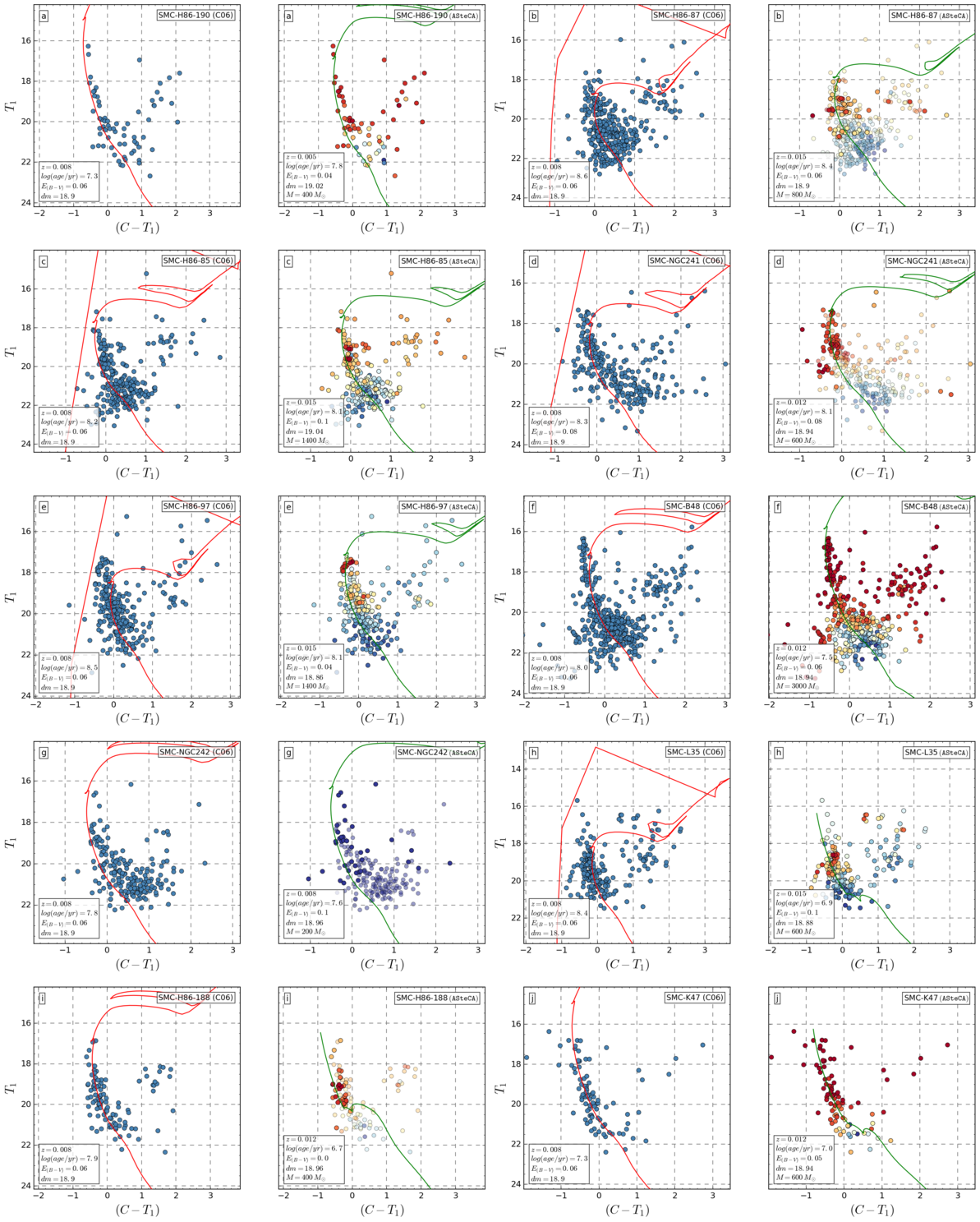}
\caption{CMDs for the C06 database.}
\label{fig:DBs_C06_2}
\end{figure*}

\begin{figure}
\includegraphics[width=\columnwidth]{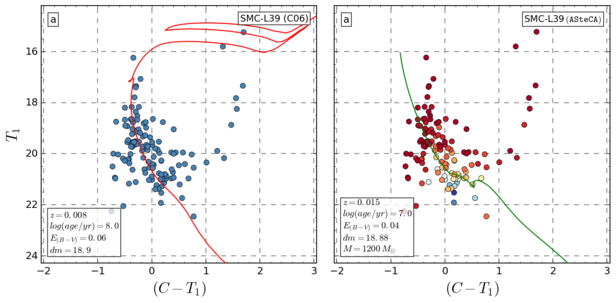}
\caption{CMDs for the C06 database.}
\label{fig:DBs_C06_3}
\end{figure}

\begin{figure*}
\includegraphics[width=2.\columnwidth]{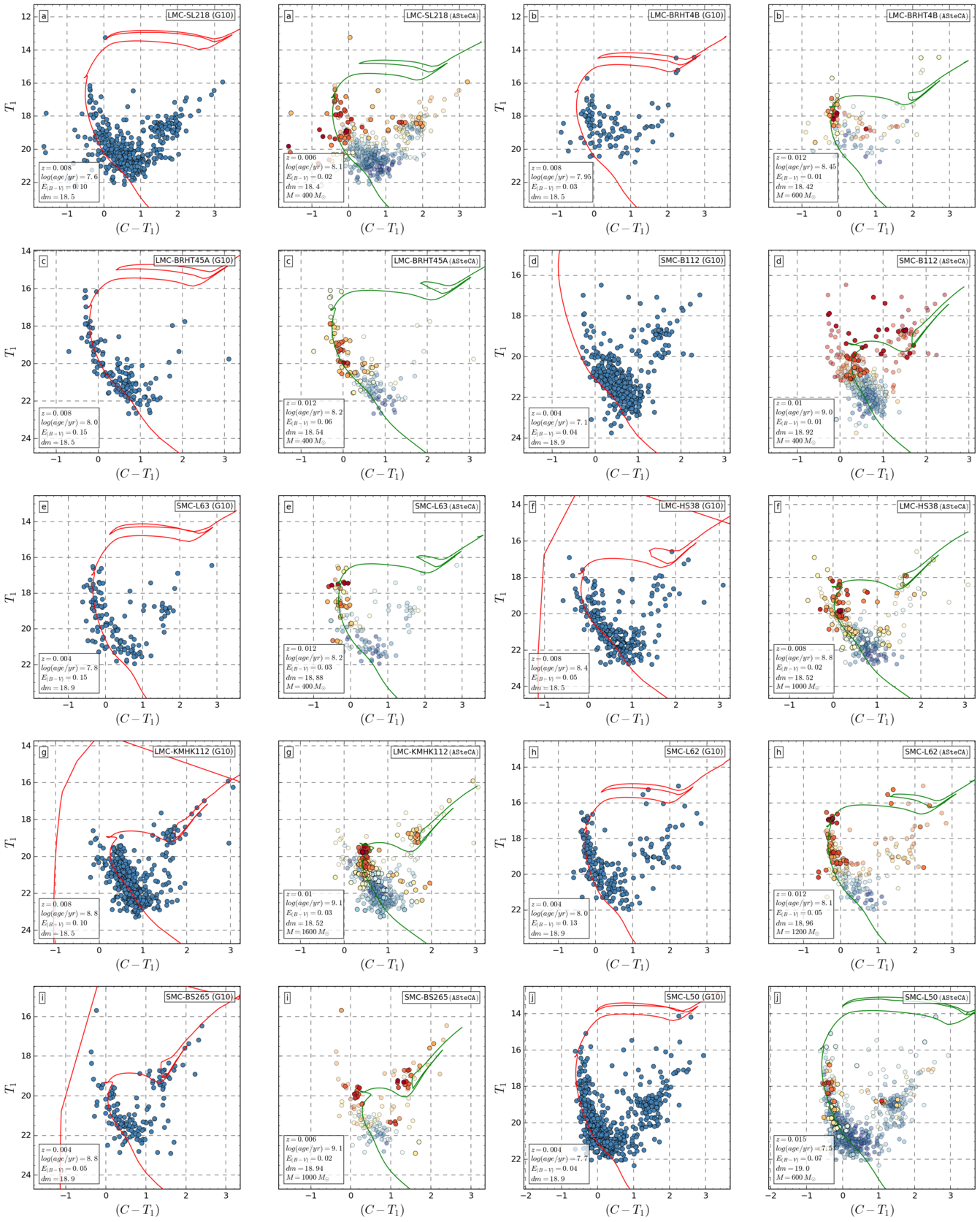}
\caption{CMDs for the G10 database.}
\label{fig:DBs_G10_0}
\end{figure*}

\begin{figure*}
\includegraphics[width=2.\columnwidth]{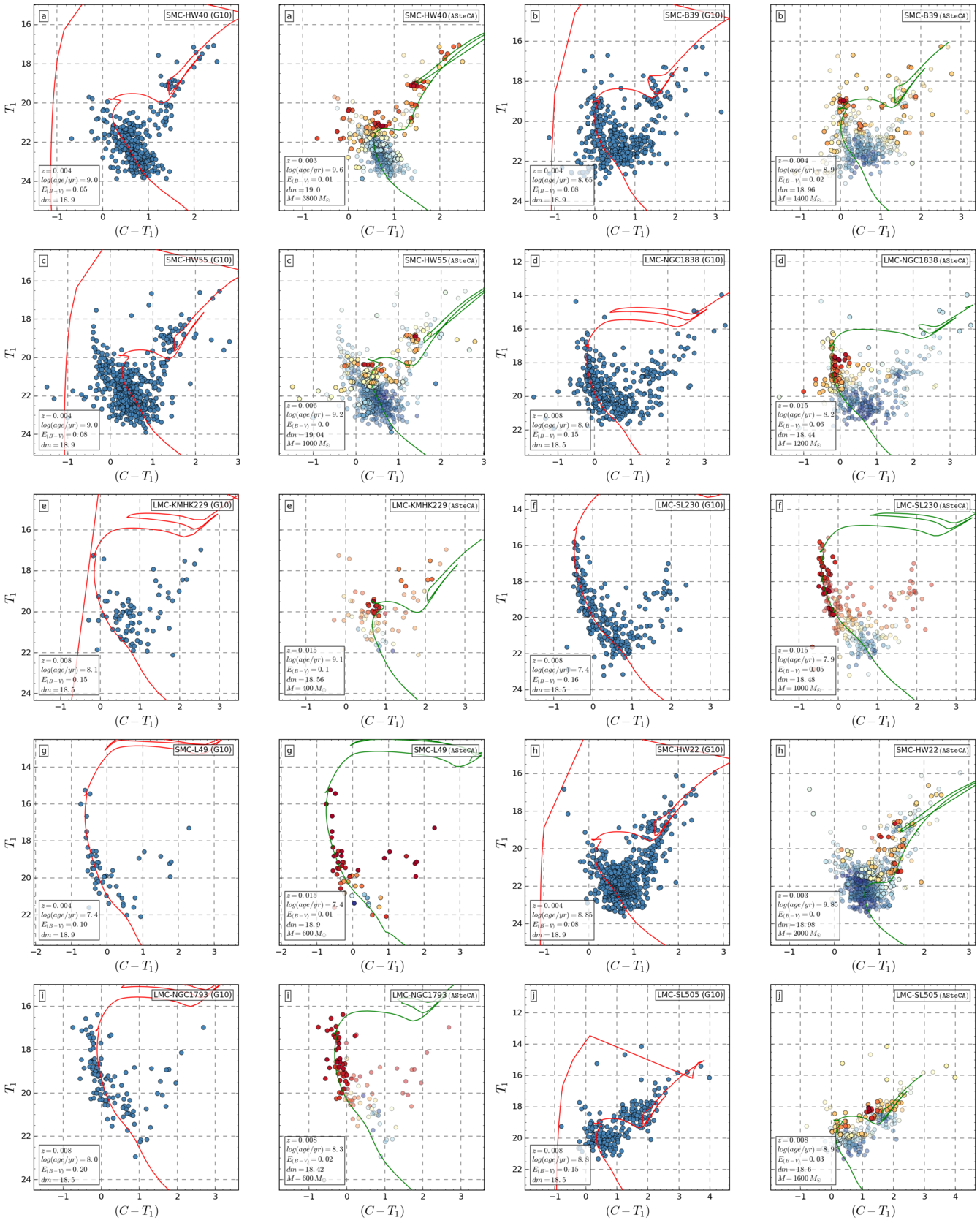}
\caption{CMDs for the G10 database.}
\label{fig:DBs_G10_1}
\end{figure*}

\begin{figure*}
\includegraphics[width=2.\columnwidth]{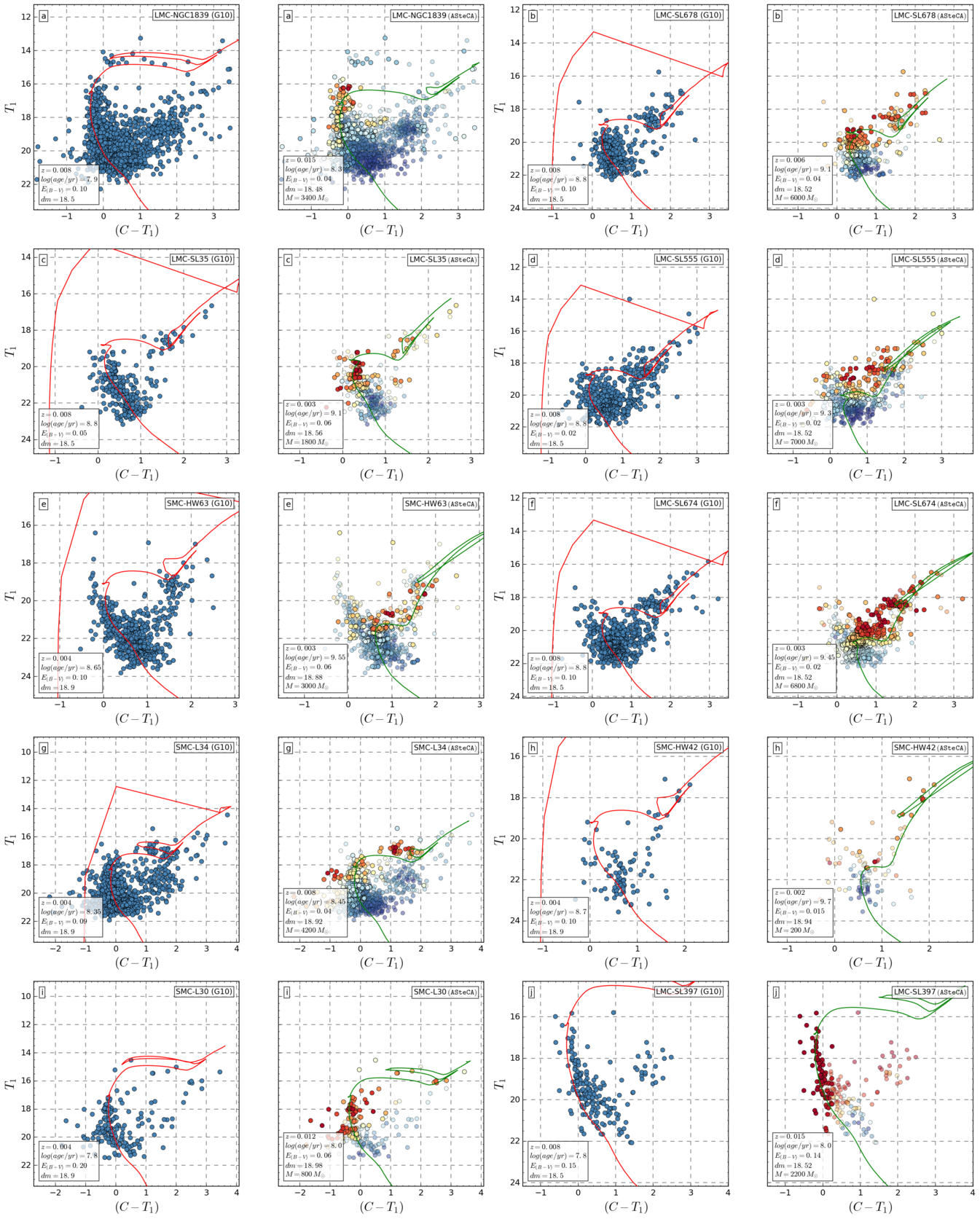}
\caption{CMDs for the G10 database.}
\label{fig:DBs_G10_2}
\end{figure*}

\begin{figure*}
\includegraphics[width=2.\columnwidth]{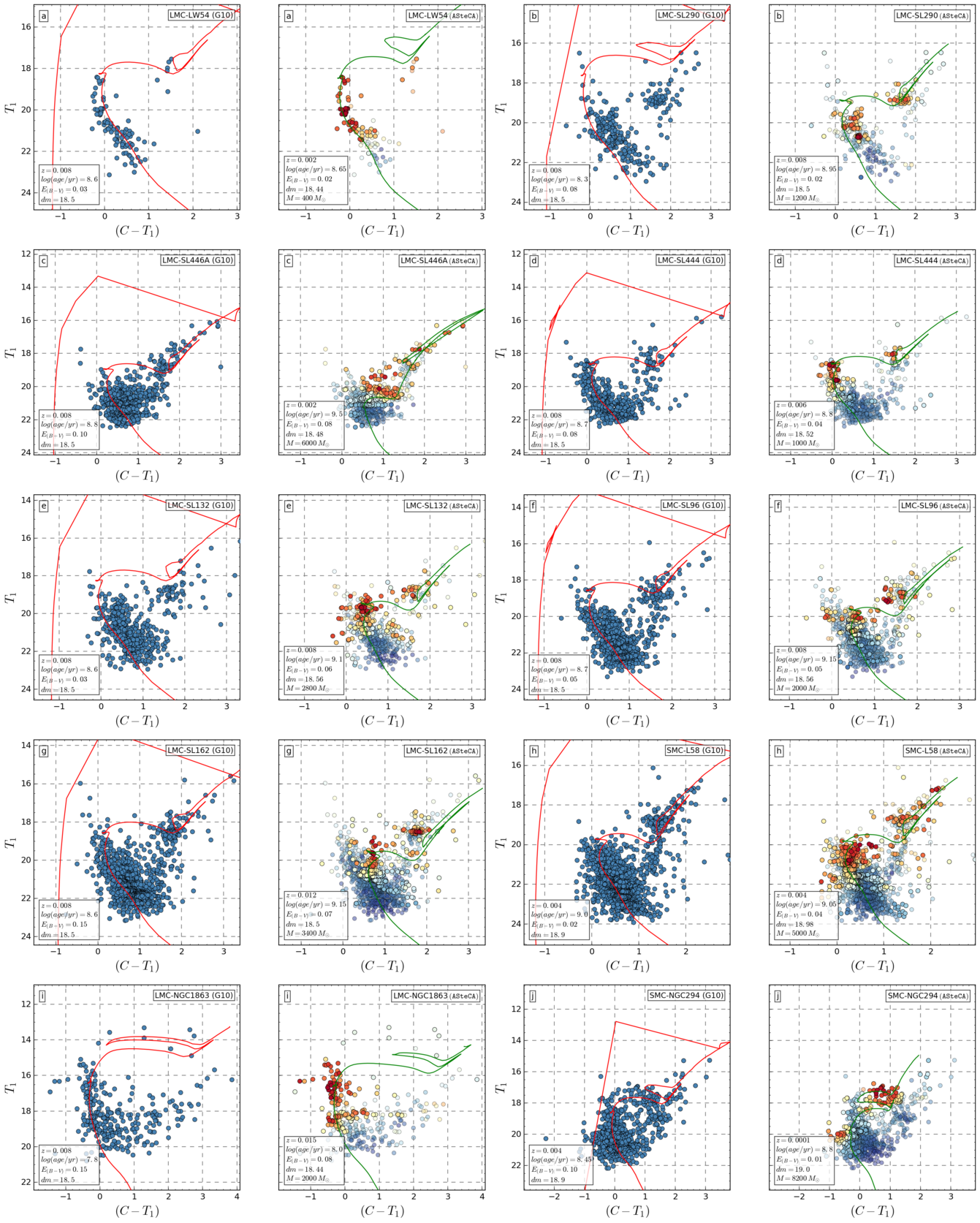}
\caption{CMDs for the G10 database.}
\label{fig:DBs_G10_3}
\end{figure*}

\begin{figure*}
\includegraphics[width=2.\columnwidth]{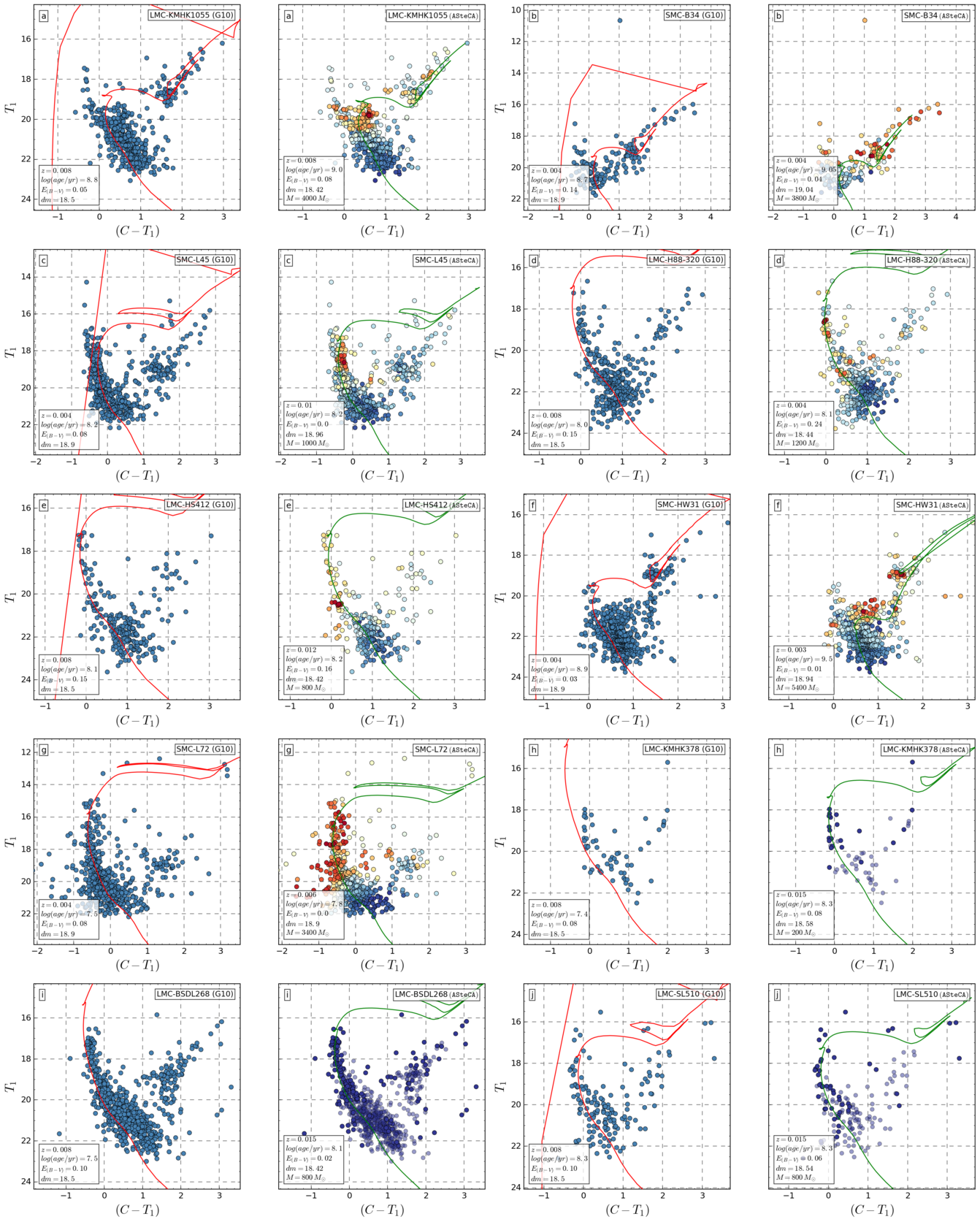}
\caption{CMDs for the G10 database.}
\label{fig:DBs_G10_4}
\end{figure*}

\begin{figure*}
\includegraphics[width=2.\columnwidth]{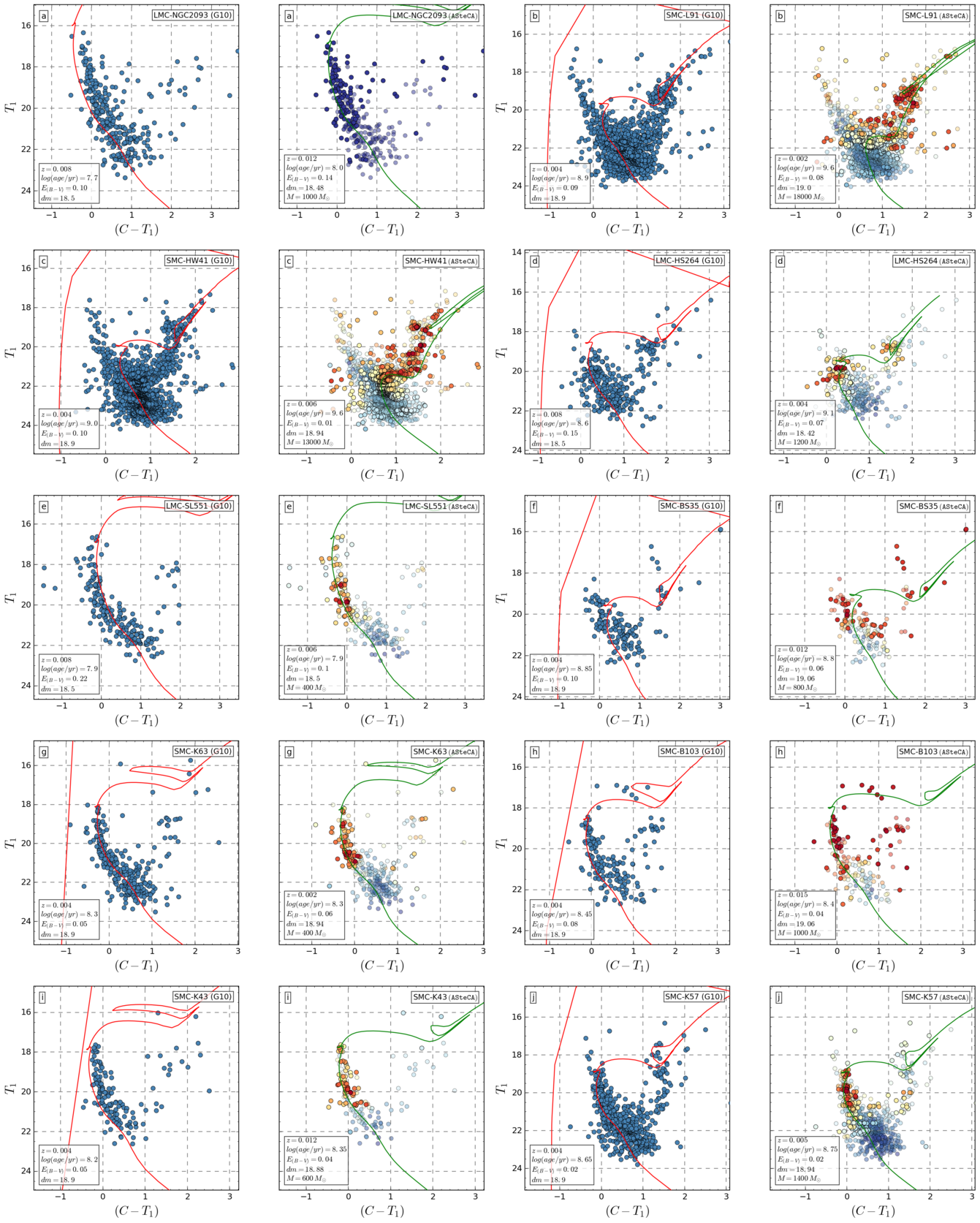}
\caption{CMDs for the G10 database.}
\label{fig:DBs_G10_5}
\end{figure*}

\begin{figure*}
\includegraphics[width=2.\columnwidth]{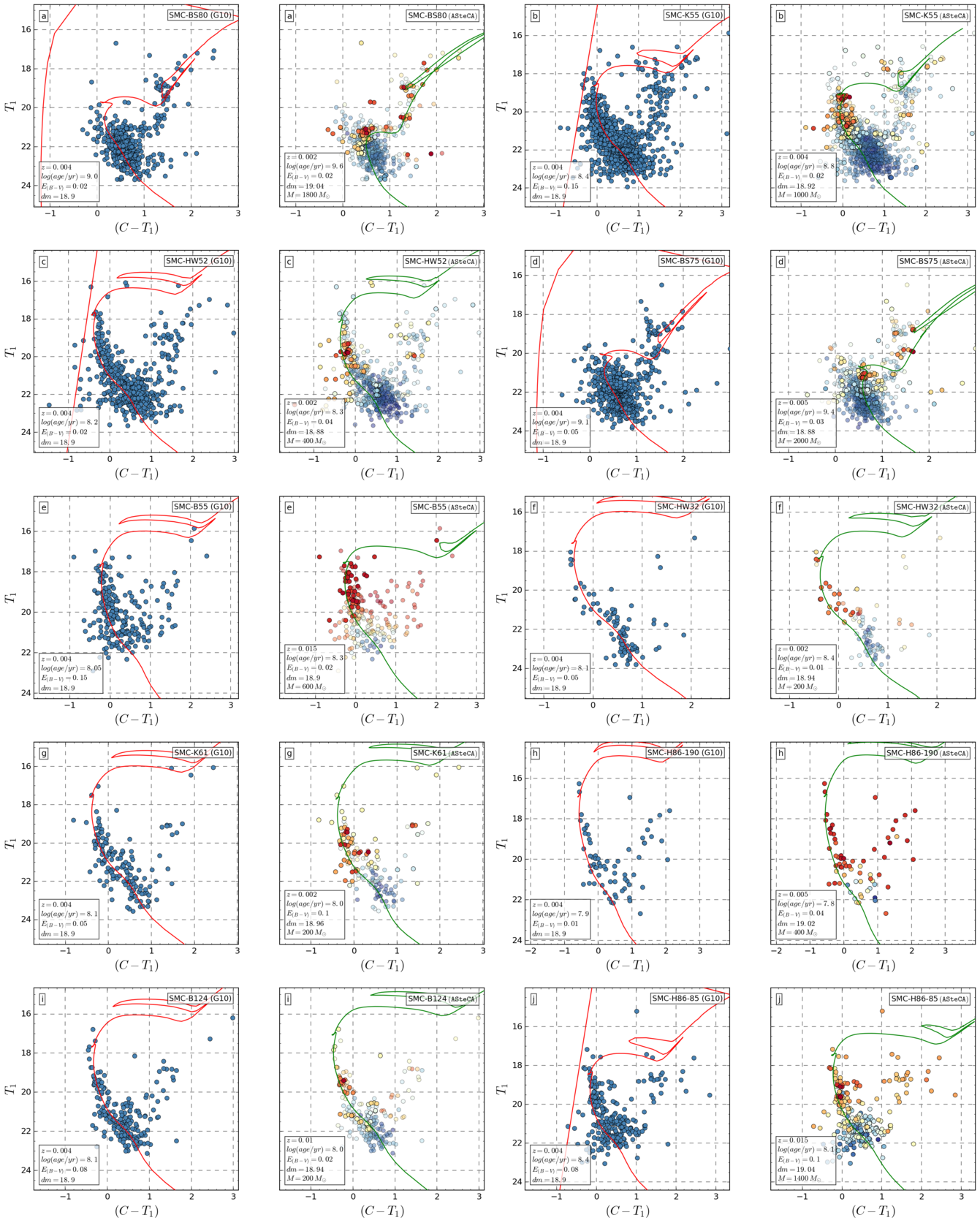}
\caption{CMDs for the G10 database.}
\label{fig:DBs_G10_6}
\end{figure*}

\begin{figure*}
\includegraphics[width=2.\columnwidth]{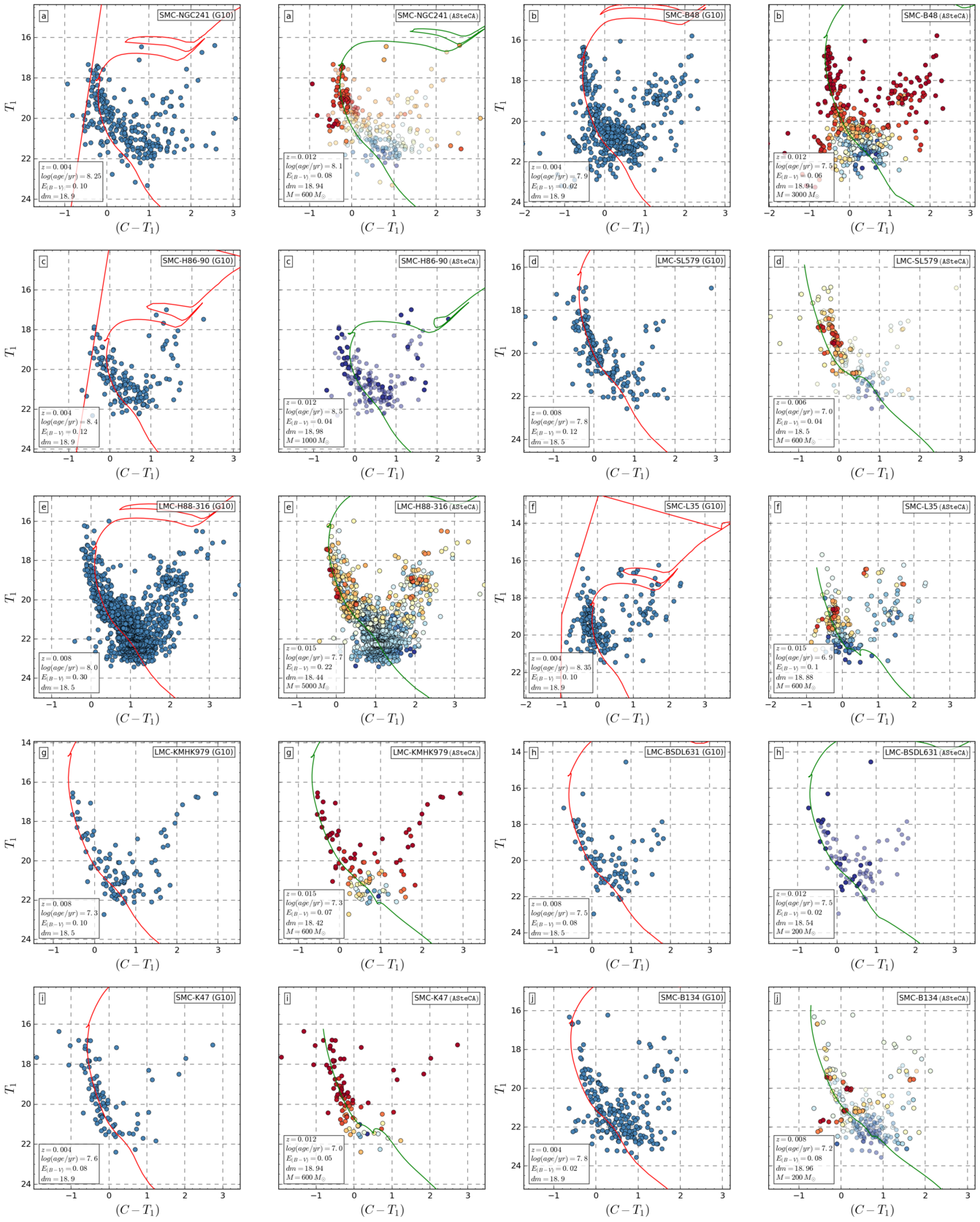}
\caption{CMDs for the G10 database.}
\label{fig:DBs_G10_7}
\end{figure*}

\begin{figure}
\centering
\includegraphics[width=\hsize]{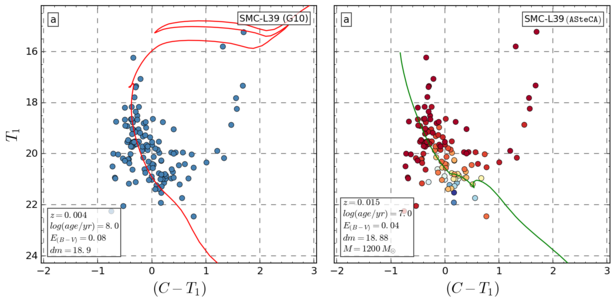}
\caption{CMDs for the G10 database.}
\label{fig:DBs_G10_8}
\end{figure}

Figs.~\ref{fig:DBs_P99_0} to~\ref{fig:DBs_G10_8} present the CMDs of clusters
cross-matched with our own sample in the databases P99, P00, C06, and G10,
i.e.:\ those that used the isochrone fit method in their analysis.
Same data distribution in the plotted CMDs as that described for
Fig.~\ref{fig:outliers0}.

\clearpage

\clearpage

\clearpage

\clearpage

\clearpage

\clearpage

\clearpage

\clearpage

\clearpage

\clearpage

\clearpage

\clearpage

\clearpage

\clearpage

\clearpage

\clearpage

\clearpage

\section{Mathematical description of the age-metallicity relationship}
\label{apdx:amr_description}

We use the KDE technique described in Sect.~\ref{sec:param-dist} to
generate an AMR representative of the observed data, with some important
improvements over previous methods.
First, unlike a regular histogram, a Gaussian density map has no dependence on
the number, size (fixed or variable) or location of bins.
Second, the errors in the two parameters used to obtain the density map (age and
metallicity), are organically included in the function that generates it
(as explained in Sect.~\ref{sec:param-dist}). No ad-hoc procedure is needed to
incorporate the information carried by these values, into the final AMR.\@

The process of creating an AMR function
requires assigning a unique [Fe/H] to a single age value, for the available age
range.
A dense grid is created to divide the age-metallicity 2D density map into N
steps of 0.01 dex width, covering the ranges of both parameters. Every point in
this grid is evaluated in the KDE map and its value ($w_{i}$) is stored, along
with its age-metallicity coordinates ($age_{i},\;[\mathrm{Fe/H}]_{i}$).
The N ages in the grid are then associated to N single representative [Fe/H]
values, obtained as the mean metallicity value weighted by the KDE function at
that particular age. The formal equation can be written as

\begin{equation}
\overline{[Fe/H]}_{age_i}=\frac{\sum w_i {[Fe/H]}_i}{\sum w_i}
\label{eq:w-feh}
\end{equation}

\noindent where the summations are performed over the N steps in the metallicity
range, ${[Fe/H]}_i$ is the metallicity value at step $i$, and $w_i$ is the value
of the 2D KDE map for that fixed age and metallicity coordinates.
The $age_i$ subindex in Eq.~\ref{eq:w-feh} indicates that this mean
metallicity was calculated for a fixed age value, and thus represents a unique
point in the AMR.\@
A similar version of this method was employed  in~\citet[][see Eq. 3]{Noel_2009}
to derive AMR estimates for three observed fields.
We apply the above formula to all ages in the grid defined at the
beginning of the process. The standard deviation for each $\overline{[Fe/H]}_
{age_i}$ value is calculated as

\begin{equation}
\sigma_{age_i}^2=\frac{\sum w_i \sum [w_i {({[Fe/H]}_i -
\overline{[Fe/H]}_{age_i})}^2]}{{(\sum w_i)}^2 - \sum w_i^2}
\label{eq:w-std-dev}
\end{equation}

\noindent where again all summations are applied over N, and the descriptions
given for the parameters in Eq.~\ref{eq:w-feh} apply.
At this point, this method already gives us an AMR function estimate, since
every age step is mapped to a unique metallicity. The downsides are that
the AMR is noisy due to the very small step of 0.01 dex used, and the
associated errors are quite large.
This latter effect arises because the weighted standard deviation, Eq.~\ref
{eq:w-std-dev}, is affected not only by errors in both measured
parameters but also by the intrinsic dispersion in the metallicity values found
for any given age.
We therefore calculate the average [Fe/H] for an age interval, rather
than assigning a metallicity value to each age step in the grid. Dividing the
age range into intervals requires a decision about the step width, much like
when constructing a histogram, bringing back the issue of binning.
We have two advantages here: a) we use Knuth's algorithm (see
Sect.~\ref{ssec:synth-match}) to obtain the optimal binning for our data, and
b) the final AMR function is very robust to changes in the binning method
selected, so even the previous choice is not crucial in determining the
shape of our AMR.\@
Finally, the $\overline{[Fe/H]}_{age_i}$ values obtained for every ${age_i}$
within a defined age interval, are averaged. Errors are propagated through the
standard formula, disregarding covariant terms~\citep[][Eq. 3.14]
{Bevington_2003}.\\



\begin{figure}
\centering
\includegraphics[width=\hsize]{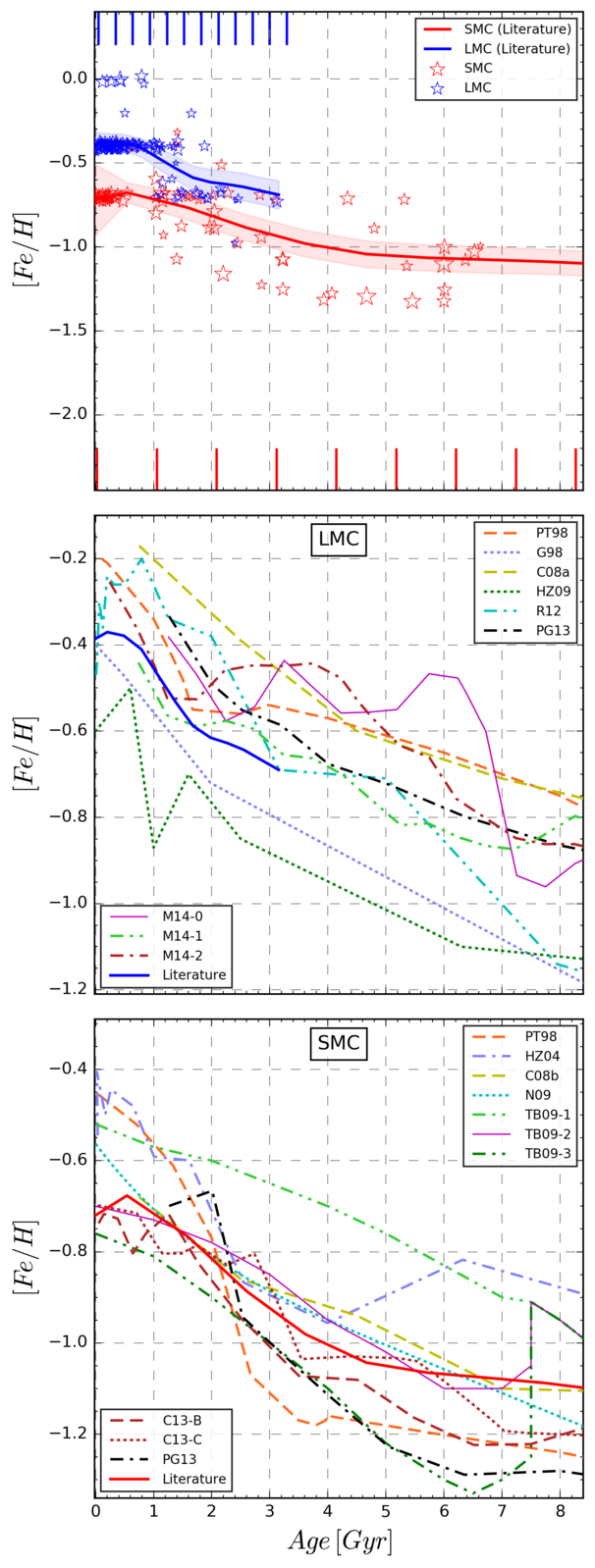}
\caption{Age-metallicity relationships for our set of 239 clusters, using
$\log\mathrm{(age/yr)}$ and [Fe/H] values taken from the literature.}
\label{fig:amr_lit}
\end{figure}

In Fig.~\ref{fig:amr_lit} we show the AMRs for both Clouds, generated using
metallicity and age values taken from the ``literature'' articles, see
Table.~\ref{tab:literature}.
Most clusters in these works are assigned fixed metallicities of -0.7 dex (SMC)
and -0.4 dex (LMC), particularly for estimated ages below 1 Gyr. This explains
the average difference of ${\sim}0.2$ dex that can be appreciated, when compared
to \texttt{ASteCA}'s AMRs (Fig.~\ref{fig:amr}).

\end{appendix} 

\end{document}